\documentclass[sigconf]{acmart} 

\AtBeginDocument{%
  }

\renewcommand\footnotetextcopyrightpermission[1]{}

\usepackage{balance}
\usepackage{flushend}
\usepackage{graphicx}
\usepackage{subcaption}
\usepackage{algorithm}
\usepackage{algpseudocode}
\usepackage{amsmath}
\usepackage{bbm}
\usepackage{amsthm}
\usepackage{amsfonts}
\usepackage{stfloats}
\usepackage{url}
\usepackage{xspace}
\usepackage{multirow}
\usepackage{booktabs}
\usepackage{mathtools}
\usepackage{caption}
\usepackage{hyperref} 
\usepackage{color}

\usepackage{multirow}
\usepackage{bbding}
\usepackage{colortbl}
\usepackage{textcomp}
\usepackage{comment}
\usepackage{enumerate}
\usepackage{tcolorbox}
\usepackage{pifont}
\usepackage{enumitem}
\setlist[itemize]{leftmargin=*}
\usepackage{tabularx}

\usepackage{array}
\usepackage{ragged2e}
\usepackage{longtable}
\usepackage{float}

\newcolumntype{C}[1]{>{\centering\arraybackslash}p{#1}} 
\newcolumntype{Y}{>{\centering\arraybackslash}X}        
\newenvironment{DIFnomarkup}{}{}

\renewcommand{\arraystretch}{1.05}
\setlist[itemize]{leftmargin=*}

\usepackage{xcolor}
\hypersetup{
  colorlinks,
  linkcolor={blue!70!green},
  citecolor={green!70!blue},
  urlcolor={orange!70!red}
}

\usepackage{hyperref}
\makeatletter
\def\UrlAlphabet{%
      \do\a\do\b\do\c\do\d\do\e\do\f\do\g\do\h\do\i\do\j%
      \do\k\do\l\do\m\do\n\do\o\do\p\do\q\do\r\do\s\do\t%
      \do\u\do\v\do\w\do\x\do\y\do\z\do\A\do\B\do\C\do\D%
      \do\E\do\F\do\G\do\H\do\I\do\J\do\K\do\L\do\M\do\N%
      \do\O\do\P\do\Q\do\R\do\S\do\T\do\U\do\V\do\W\do\X%
      \do\Y\do\Z}
\def\UrlDigits{\do\1\do\2\do\3\do\4\do\5\do\6\do\7\do\8\do\9\do\0}
\g@addto@macro{\UrlBreaks}{\UrlOrds}
\g@addto@macro{\UrlBreaks}{\UrlAlphabet}
\g@addto@macro{\UrlBreaks}{\UrlDigits}
\makeatother

\definecolor{mygray}{gray}{.9}
\newcommand{\etal}{\textit{et al.}\xspace}
\newcommand{\ie}{\textit{i.e.}\xspace}
\newcommand{\eg}{\textit{e.g.}\xspace}

\newcommand{\mypara}[1]{\smallskip\noindent\textbf{#1.} \xspace}

\usepackage{soul}
\newcommand{\sysname}{NotaryMark\xspace}
\newcommand{\hltext}[1]{%
  \begingroup
  \def\xspace{}%
  \edef\temp{\endgroup
    \noexpand\hl{#1}}%
  \temp
}

\definecolor{revision}{RGB}{0,0,255}

\newcommand{\revstart}{\begin{color}{revision}}
\newcommand{\revend}{~\!\!\end{color}}

\newcommand{\BAimgMin}{30.43}
\newcommand{\BAimgMax}{100.00}

\usepackage[table]{xcolor}
\usepackage{pgf}

\definecolor{MorandiRed}{HTML}{eb8386}   
\definecolor{MorandiBlue}{HTML}{7EB2E6} 
\definecolor{MorandiGray}{HTML}{f5f5f5}

\definecolor{MacaRed}{HTML}{eb8386}   
\definecolor{MacaBlue}{HTML}{7EB2E6} 
\definecolor{MacaGray}{HTML}{ffffff}  
\definecolor{MacaGreen}{HTML}{a8d8b9}  
\definecolor{MacaYellow}{HTML}{f9e4ad} 
\definecolor{MacaPink}{HTML}{FFD1DC}   

\definecolor{mylightblue}{RGB}{211, 231, 248} 
\definecolor{mylightgray}{RGB}{245, 245, 245} 

\newcommand{\PCT}{0}
\newcommand{\subPCT}{0}

\newcommand{\ClampInt}[1]{%
  \pgfmathtruncatemacro{\localPCT}{max(0,min(100,#1))}%
  \global\let\PCT\localPCT
}

\newcommand{\heatBAimg}[1]{%
  \pgfmathsetmacro{\val}{100*(#1-\BAimgMin)/(\BAimgMax-\BAimgMin)}%
  \ClampInt{\val}%
  \ifnum\PCT<50
    \pgfmathtruncatemacro{\localSubPCT}{100 - (\PCT * 2)}%
    \global\let\subPCT\localSubPCT 
    \cellcolor{MacaRed!\subPCT!MacaGray}%
  \else
    \pgfmathtruncatemacro{\localSubPCT}{(\PCT - 50) * 2}%
    \global\let\subPCT\localSubPCT 
    \cellcolor{MacaBlue!\subPCT!MacaGray}%
  \fi
  \strut #1%
}

\newcommand{\BAimgMinn}{45.00}
\newcommand{\BAimgMaxn}{100.00}
\newcommand{\heatBAimgn}[1]{%
  \pgfmathsetmacro{\val}{100*(#1-\BAimgMinn)/(\BAimgMaxn-\BAimgMinn)}%
  \ClampInt{\val}%
  \ifnum\PCT<50
    \pgfmathtruncatemacro{\localSubPCT}{100 - (\PCT * 2)}%
    \global\let\subPCT\localSubPCT
    \cellcolor{MorandiRed!\subPCT!MorandiGray}%
  \else
    \pgfmathtruncatemacro{\localSubPCT}{(\PCT - 50) * 2}%
    \global\let\subPCT\localSubPCT
    \cellcolor{MorandiBlue!\subPCT!MorandiGray}%
  \fi
  \strut #1%
}

\newcommand{\FPRMin}{0}
\newcommand{\FPRMax}{20}
\newcommand{\heatFPR}[1]{%
  \pgfmathsetmacro{\val}{100*(#1-\FPRMin)/(\FPRMax-\FPRMin)}%
  \ClampInt{\val}%
  \ifnum\PCT<50
    \pgfmathtruncatemacro{\localSubPCT}{100 - (\PCT * 2)}%
    \global\let\subPCT\localSubPCT
    \cellcolor{MorandiBlue!\subPCT!MorandiGray}%
  \else
    \pgfmathtruncatemacro{\localSubPCT}{(\PCT - 50) * 2}%
    \global\let\subPCT\localSubPCT
    \cellcolor{MorandiRed!\subPCT!MorandiGray}%
  \fi
  \strut #1%
}

\newcommand{\TPRMin}{70}
\newcommand{\TPRMax}{100}
\newcommand{\heatPrecFone}[1]{%
  \pgfmathsetmacro{\val}{100*(#1-\TPRMin)/(\TPRMax-\TPRMin)}%
  \ClampInt{\val}%
  \ifnum\PCT<50
    \pgfmathtruncatemacro{\localSubPCT}{100 - (\PCT * 2)}%
    \global\let\subPCT\localSubPCT 
    \cellcolor{MorandiRed!\subPCT!MorandiGray}%
  \else
    \pgfmathtruncatemacro{\localSubPCT}{(\PCT - 50) * 2}%
    \global\let\subPCT\localSubPCT 
    \cellcolor{MorandiBlue!\subPCT!MorandiGray}%
  \fi
  \strut #1%
}

\newcommand{\SSIMMin}{0.9745}
\newcommand{\SSIMMax}{0.9828}

\newcommand{\heatSSIM}[1]{%
 \pgfmathsetmacro{\val}{100*(#1-\SSIMMin)/(\SSIMMax-\SSIMMin)}%
  \ClampInt{\val}%
  \ifnum\PCT<50
    \pgfmathtruncatemacro{\localSubPCT}{100 - (\PCT * 2)}%
    \global\let\subPCT\localSubPCT
    \cellcolor{MorandiRed!\subPCT!MorandiGray}%
  \else
    \pgfmathtruncatemacro{\localSubPCT}{(\PCT - 50) * 2}%
    \global\let\subPCT\localSubPCT
    \cellcolor{MorandiBlue!\subPCT!MorandiGray}%
  \fi
  \strut #1%
}

\newcommand{\LPIPSMin}{0.0102}
\newcommand{\LPIPSMax}{0.0148}
\newcommand{\heatLPIPS}[1]{%
 \pgfmathsetmacro{\val}{100*(#1-\LPIPSMin)/(\LPIPSMax-\LPIPSMin)}%
  \ClampInt{\val}%
  \ifnum\PCT<50
    \pgfmathtruncatemacro{\localSubPCT}{100 - (\PCT * 2)}%
    \global\let\subPCT\localSubPCT
    \cellcolor{MorandiBlue!\subPCT!MorandiGray}%
  \else
    \pgfmathtruncatemacro{\localSubPCT}{(\PCT - 50) * 2}%
    \global\let\subPCT\localSubPCT
    \cellcolor{MorandiRed!\subPCT!MorandiGray}%
  \fi
  \strut #1%
}
\newcommand{\FIDMin}{74.47}
\newcommand{\FIDMax}{95.81}
\newcommand{\heatFID}[1]{%
 \pgfmathsetmacro{\val}{100*(#1-\FIDMin)/(\FIDMax-\FIDMin)}%
  \ClampInt{\val}%
  \ifnum\PCT<50
    \pgfmathtruncatemacro{\localSubPCT}{100 - (\PCT * 2)}%
    \global\let\subPCT\localSubPCT
    \cellcolor{MorandiBlue!\subPCT!MorandiGray}%
  \else
    \pgfmathtruncatemacro{\localSubPCT}{(\PCT - 50) * 2}%
    \global\let\subPCT\localSubPCT
    \cellcolor{MorandiRed!\subPCT!MorandiGray}%
  \fi
  \strut #1%
}

\newcommand{\SSIMMinn}{0.9745}
\newcommand{\SSIMMaxn}{0.9828}

\newcommand{\heatSSIMn}[1]{%
 \pgfmathsetmacro{\val}{100*(#1-\SSIMMinn)/(\SSIMMaxn-\SSIMMinn)}%
  \ClampInt{\val}%
  \ifnum\PCT<50
    \pgfmathtruncatemacro{\localSubPCT}{100 - (\PCT * 2)}%
    \global\let\subPCT\localSubPCT
    \cellcolor{MacaYellow!\subPCT!MorandiGray}%
  \else
    \pgfmathtruncatemacro{\localSubPCT}{(\PCT - 50) * 2}%
    \global\let\subPCT\localSubPCT
    \cellcolor{MacaGreen!\subPCT!MorandiGray}%
  \fi
  \strut #1%
}

\newcommand{\LPIPSMinn}{0.0102}
\newcommand{\LPIPSMaxn}{0.0148}
\newcommand{\heatLPIPSn}[1]{%
 \pgfmathsetmacro{\val}{100*(#1-\LPIPSMinn)/(\LPIPSMaxn-\LPIPSMinn)}%
  \ClampInt{\val}%
  \ifnum\PCT<50
    \pgfmathtruncatemacro{\localSubPCT}{100 - (\PCT * 2)}%
    \global\let\subPCT\localSubPCT
    \cellcolor{MacaGreen!\subPCT!MorandiGray}%
  \else
    \pgfmathtruncatemacro{\localSubPCT}{(\PCT - 50) * 2}%
    \global\let\subPCT\localSubPCT
    \cellcolor{MacaYellow!\subPCT!MorandiGray}%
  \fi
  \strut #1%
}

\newcommand{\FIDMinn}{74.47}
\newcommand{\FIDMaxn}{95.81}
\newcommand{\heatFIDn}[1]{%
 \pgfmathsetmacro{\val}{100*(#1-\FIDMinn)/(\FIDMaxn-\FIDMinn)}%
  \ClampInt{\val}%
  \ifnum\PCT<50
    \pgfmathtruncatemacro{\localSubPCT}{100 - (\PCT * 2)}%
    \global\let\subPCT\localSubPCT
    \cellcolor{MacaGreen!\subPCT!MorandiGray}%
  \else
    \pgfmathtruncatemacro{\localSubPCT}{(\PCT - 50) * 2}%
    \global\let\subPCT\localSubPCT
    \cellcolor{MacaYellow!\subPCT!MorandiGray}%
  \fi
  \strut #1%
}

\usepackage{etoolbox}
\makeatletter
\patchcmd{\hyper@makecurrent}{%
    \ifx\Hy@param\Hy@chapterstring
        \let\Hy@param\Hy@chapapp
    \fi
}{%
    \iftoggle{inappendix}{
        \@checkappendixparam{chapter}%
        \@checkappendixparam{section}%
        \@checkappendixparam{subsection}%
        \@checkappendixparam{subsubsection}%
        \@checkappendixparam{paragraph}%
        \@checkappendixparam{subparagraph}%
    }{}%
}{}{\errmessage{failed to patch}}

\newcommand*{\@checkappendixparam}[1]{%
    \def\@checkappendixparamtmp{#1}%
    \ifx\Hy@param\@checkappendixparamtmp
        \let\Hy@param\Hy@appendixstring
    \fi
}
\makeatletter

\newtoggle{inappendix}
\togglefalse{inappendix}

\apptocmd{\appendix}{\toggletrue{inappendix}}{}{\errmessage{failed to patch}}

\usepackage{tikz}
\usepackage{oplotsymbl}

\tcbuselibrary{breakable}

\begin{document}
\pagestyle{plain}
\author{Xinyi Yu}
\affiliation{%
  \institution{Xi'an Jiaotong University}
  \city{Xi'an}
  \country{China}
}
\email{xinyiyu@stu.xjtu.edu.cn}

\author{Linkang Du}
\authornote{Corresponding authors.}
\affiliation{%
  \institution{Xi'an Jiaotong University}
  \city{Xi'an}
  \country{China}
}
\email{linkangd@xjtu.edu.cn}

\author{Zhou Su}
\authornotemark[2]
\affiliation{%
  \institution{Xi'an Jiaotong University}
  \city{Xi'an}
  \country{China}
}
\email{zhousu@ieee.org}

\author{Min Chen}
\affiliation{%
  \institution{Vrije Universiteit Amsterdam}
  \city{Amsterdam}
  \country{The Netherlands}
}
\email{m.chen2@vu.nl}

\author{Yuntao Wang}
\affiliation{%
  \institution{Xi'an Jiaotong University}
  \city{Xi'an}
  \country{China}
}
\email{yuntao.wang@xjtu.edu.cn}

\author{Zhikun Zhang}
\affiliation{%
  \institution{Zhejiang University}
  \city{Hangzhou}
  \country{China}
}
\email{zhikun@zju.edu.cn}

\title{\sysname: Multi-Bit Watermarking for Black-Box Dataset Usage Auditing in Text-to-Image Diffusion Models}



\begin{abstract}
As generative artificial intelligence (AI) models increasingly require data, more publicly available yet licensed data is being harvested and used without authorization, \eg, Disney accuses Google of using AI to engage in copyright infringement on a ``massive scale.'' 
Although several previous studies have proposed watermarking methods for dataset copyright auditing, their applicability to the text-to-image setting remains notably limited. 
The core drawback is that watermarks embedded by existing techniques are often treated as noise and are gradually filtered out and attenuated during the latent noising-denoising process of diffusion model training.

We propose \sysname, to our knowledge the first multi-bit watermarking method specifically designed for black-box auditing of T2I training dataset use.
The core idea of \sysname is to align the watermark with the latent denoising of diffusion models, thereby overcoming the signal decay that plagues existing multi-bit methods during sampling.
We evaluate \sysname across three datasets and three diffusion models spanning watermark capacities from 8 to 1024 bits.
With a watermark capacity of below 32 bits, it achieves a high average bit accuracy of 0.85-0.90.
With a watermark capacity exceeding 32 bits, it maintains bit accuracy between 0.67 and 0.80, consistently outperforming existing multi-bit watermark methods whose bit accuracy is around 0.42-0.60. 
Furthermore, we design a voting-based aggregation scheme to boost the bit accuracy with a modest budget of 30 queries.
Building on this, we incorporate error-correcting codes to enable reliable decoding and tracing of the embedded identifier information.
Robustness evaluation against fifteen evasion attacks spanning seven categories shows that \sysname remains robust under most evaluated settings, while stronger attacks generally suppress watermark evidence only at the cost of substantial utility degradation, demonstrating its practical reliability against malicious tampering.

\end{abstract}

\ccsdesc[500]{Security and privacy~Digital rights management}
\ccsdesc[500]{Computing methodologies~Artificial intelligence}

\keywords{Dataset Watermarking, Black-box Auditing, Text-to-Image Diffusion Models.} 

\maketitle

\section{Introduction}
Generative artificial intelligence (AI) has advanced rapidly in recent years and is now widely used in areas such as content creation~\cite{dehouche2023s,zhang2025creatidesign,gao2026multi} and industrial design~\cite{barros2024designing,tao2025designweaver,jiang2024haigen,geng2025diffdesign}. 
Among the various applications, text-to-image (T2I) diffusion models have become one of the most prominent paradigms.
Their strong generative performance relies on large-scale, high-quality training datasets. 
However, as more datasets become readily available, it has become much easier to collect them without permission and repurpose them for new uses, raising serious concerns about data copyright~\cite{longpre2024large}.
A recent example occurred in December 2025, when Disney issued a cease-and-desist letter to Google, alleging that Google's generative AI facilitates copyright infringement at a ``massive scale'' across both training and distribution pipelines~\cite{axios_disney_google_2025}.
This high-profile legal tension underscores the urgent need for reliable model-auditing mechanisms that can provide reproducible and technical evidence for dataset copyright verification~\cite{atli2022effectiveness,du2025dataset,huang2024general,song2019auditing,pang2023black}.

Dataset watermarking has emerged as a promising solution for copyright protection.
The key principle involves embedding verifiable identifiers into training samples, ensuring these marks propagate into the model outputs generated by models trained on them~\cite{li2025towards}.
This enables dataset owners to audit potential misuse through black-box queries and obtain evidence.
These methods can be categorized into zero-bit and multi-bit schemes based on watermark capacity~\cite{li2021survey,zhong2023brief}.
While zero-bit methods provide only a binary verdict on whether the dataset is misused, multi-bit methods offer significantly greater practical utility by carrying metadata, such as dataset IDs, timestamps, or licensee fingerprints~\cite{xing2025optmark,li2025hmark,qu2025provably}.
Such high-capacity information facilitates fine-grained provenance tracing and responsibility attribution, essential for multi-party license management and real-world legal enforcement~\cite{fernandez2023stable}.

\mypara{Challenges}
Despite its potential, stably injecting multi-bit watermarks into the training pipeline of T2I diffusion models remains non-trivial. 
On one hand, traditional multi-bit watermarking schemes~\cite{zhang2019robust,fernandez2022watermarking} are primarily designed for robustness against media post-processing operations, rather than being optimized to handle watermark signal distortion caused by noise perturbation, iterative denoising, and latent-space modeling during the training and inference of T2I diffusion models. 
On the other hand, emerging methods tailored for diffusion models~\cite{cui2023diffusionshield,li2025hmark} are often limited to pixel-domain or unconditional generation, failing to withstand the complex cross-attention and latent-space learning inherent in modern T2I architectures. 
As a result, the watermark signals generated by these methods are prone to distortion during T2I model training and fine-tuning.
Empirically, we observe that in practical T2I settings, prior watermarking methods often struggle to achieve reliable bit recovery, with bit accuracy rarely exceeding 60\% and frequently approaching random guessing.
Thus, to make multi-bit watermarking viable for copyright verification, three challenges should be addressed: 
1) how to ensure that the watermark signal survives stochastic model training and fine-tuning, and is persistently learned by the downstream model; 
2) how to reliably separate and recover the watermark signal from the complex semantic content of generated images, which may differ substantially from those of the protected training samples; and 
3) how to reconstruct a stable and high-capacity provenance message despite the inherent randomness of diffusion sampling.

\mypara{Our Proposal}
To address these challenges, we propose \sysname, to our knowledge the first multi-bit watermarking framework specifically designed for black-box auditing of T2I training-dataset use. 
During watermark injection, \sysname represents the message as a structured residual in the diffusion-aligned latent space, improving its persistence across the indirect data-model-image channel. 
For verification, we introduce a U-Net watermark predictor supervised by the original residual and formulate extraction as generative denoising, enabling watermark recovery from generated latents despite semantic prompt shifts and image-content interference. 
Finally, \sysname aggregates complementary bit evidence from multiple generated images and applies BCH error correction to mitigate stochastic decoding errors and recover a reliable provenance message. 
A joint objective further balances watermark predictability, decodability, and visual imperceptibility.

\mypara{Evaluation}
To assess the effectiveness of \sysname, we perform extensive experiments across multiple datasets, including Pok\'{e}mon~\cite{pinkney2022pokemon}, CelebA~\cite{liu2015deep}, and WikiArt~\cite{saleh2015large}, and across different model architectures, including Stable Diffusion v1.5~\cite{rombach2022high}, Stable Diffusion XL~\cite{podell2023sdxl}, and Kandinsky 2.2~\cite{razzhigaev2023kandinsky}.
Empirical results demonstrate the exceptional scalability of \sysname across a range of watermark capacities. 
\sysname achieves a bit accuracy of 89.50\% at 8 bits, and maintains a competitive accuracy of 67.04\% even at an ultra-high capacity of 1024 bits. 
Compared with existing multi-bit watermarking schemes, \sysname exhibits a substantial performance margin in complex environments. 
Notably, in black-box auditing scenarios, our proposed voting-based aggregation method achieves 100\% recovery of watermark information with a minimal budget of 30 queries at capacities up to 32 bits, and further integrates error-correcting codes to support exact identifier recovery.
Furthermore, \sysname maintains bit accuracy of around 85\% when standard image-processing operations, such as JPEG compression, resizing, Gaussian blur, hue adjustment, and sharpness adjustment, are applied either as preprocessing to the protected dataset or as post-processing to generated outputs.
It also demonstrates robustness across adaptive evasion attacks, including VAE re-encoding, image regeneration, text re-captioning, watermarked-data filtering, and continued fine-tuning.

\mypara{Contributions}
Our main contributions are as follows:
\begin{itemize}
  \item We propose a novel dataset watermarking framework that is intrinsically aligned with the latent space of diffusion models.
  \sysname can embed multi-bit watermarks into a latent space of diffusion models, preventing the signal from being treated as noise and filtered out during training, thereby enabling more stable recovery of copyright messages under black-box auditing.
  \item We design a joint optimization strategy to balance the watermark's predictability, decodability, and imperceptibility. 
  We further employ stage-wise optimization to ensure reliable watermark learning and verification in high-capacity settings.
  \item We conduct a comprehensive evaluation in black-box T2I auditing settings and introduce a multi-sample voting scheme to mitigate diffusion stochasticity and improve recovery stability.
  \item We build a complete end-to-end tracing pipeline that incorporates error-correcting codes. This extends the evaluations in existing work, which are limited to bit-level recovery.
  \sysname achieves over 99\% tracing success rate for watermark lengths up to 256 bits.
  
  
\end{itemize}

\section{Preliminaries}

\subsection{Text-to-Image Diffusion Models}
\label{sec:prelim_t2i_diffusion}

Text-to-image generation has progressed from GANs~\cite{zhang2017stackgan} and autoregressive models~\cite{ramesh2021zero} to diffusion models~\cite{rombach2022high}. 
Unlike GANs, diffusion models synthesize images via a gradual noising-and-denoising procedure, which is more stable and produces high-fidelity results~\cite{ho2020denoising}. 
Latent diffusion models further move the diffusion process from pixel space to a compact latent space using a VAE~\cite{kingma2013auto}, forming the foundation of Stable Diffusion and related systems~\cite{rombach2022high}. 
Below, we briefly introduce its core modules.

\mypara{Forward Diffusion}
Latent Diffusion Models first map an image $x_0$ to a compact latent representation $z_0=\mathcal{E}(x_0)$ using a VAE encoder, where $\mathcal{E}$ denotes the VAE encoder, and perform diffusion in the latent space. They define a fixed forward Markov process that progressively perturbs $z_0$ with Gaussian noise under a schedule $\{\beta_t\}_{t=1}^T$, where $T$ is the total number of diffusion steps and $\beta_t\in(0,1)$ controls the noise variance added at step $t$. Let $\alpha_t = 1-\beta_t$ denote the signal-retention coefficient at step $t$ and $\bar{\alpha}_t=\prod_{s=1}^t \alpha_s$ denote the cumulative signal-retention coefficient up to step $t$. 
The noisy latent at step $t$ has a closed form:
\begin{align}
z_t = \sqrt{\bar{\alpha}_t}\,z_0 + \sqrt{1-\bar{\alpha}_t}\,\epsilon,\quad \epsilon\sim\mathcal{N}(0,\mathbf{I}).
\end{align}
Where $\epsilon$ denotes standard Gaussian noise and $\mathbf{I}$ is the identity covariance matrix. As $t$ increases, the latent signal gradually vanishes and $z_t$ approaches isotropic Gaussian noise.

\mypara{Reverse Diffusion}
The generative model learns a reverse process that denoises $z_t$ toward $z_{t-1}$, conditioned on text embedding $c$. In practice, a U-Net denoiser parameterized by $\theta$ predicts the injected noise $\epsilon$ via $\epsilon_\theta(z_t,t,c)$, leading to the standard MSE objective:
\begin{align}
\mathcal{L}_{\text{diff}}
= \mathbb{E}_{x_0,\,t,\,\epsilon}\Big[
\big\|\epsilon - \epsilon_\theta(z_t,t,c)\big\|_2^2
\Big],
\label{eq:diff_loss}
\end{align}
where $z_0=\mathcal{E}(x_0)$ and $z_t$ is sampled from the forward process. At inference, generation starts from Gaussian noise $z_T\sim\mathcal{N}(0,\mathbf{I})$ and iteratively applies denoising updates for $t=T,\dots,1$ to obtain $z_0$, which is finally decoded into an image $\hat{x}=\mathcal{D}(z_0)$, where $\mathcal{D}$ denotes the VAE decoder.

\subsection{Dataset Watermark for Diffusion Pipeline}
Dataset watermarking has emerged as a proactive mechanism for detecting unauthorized use of datasets for training or fine-tuning generative models~\cite{du2025sok,li2023black}. 
It aims to inject verifiable identity signals into a dataset at the time of release.
If a third party subsequently trains a generative model (\eg, a Diffusion Model~\cite{li2025towards,ren2024entruth}, or LLM~\cite{zhang2025leave,li2025data}) on these watermarked data, the owner can verify the unauthorized usage by querying the model's outputs. 
Based on the capacity of the embedded information, existing methods can be broadly categorized into \textbf{zero-bit} and \textbf{multi-bit} watermarks.

\mypara{Zero-bit Watermark}
Zero-bit dataset watermarking methods are primarily designed to detect whether a specific dataset has been used in model training or fine-tuning, rather than to extract any explicit data message~\cite{zhong2023brief}.
Representative works such as SIREN~\cite{li2025towards}, FT-Shield~\cite{cui2025ft}, DIAGNOSIS~\cite{wang2023diagnosis}, and EnTruth~\cite{ren2024entruth} typically inject lightweight, imperceptible, yet learnable patterns into the released samples. 
During auditing, the dataset owner queries the suspect model with specific prompts to detect whether the generated outputs exhibit predefined behaviors or statistically significant biases.

While these methods are effective for simple ownership verification, their binary nature (Yes or No) inherently limits their utility. 
They cannot carry fine-grained information required for complex scenarios, such as distinguishing among multiple downstream licensees or performing detailed accountability attribution~\cite{min2024watermark,zhu2024flexible}.

\begin{figure}[t]
  \centering
  \includegraphics[width=0.9\columnwidth]{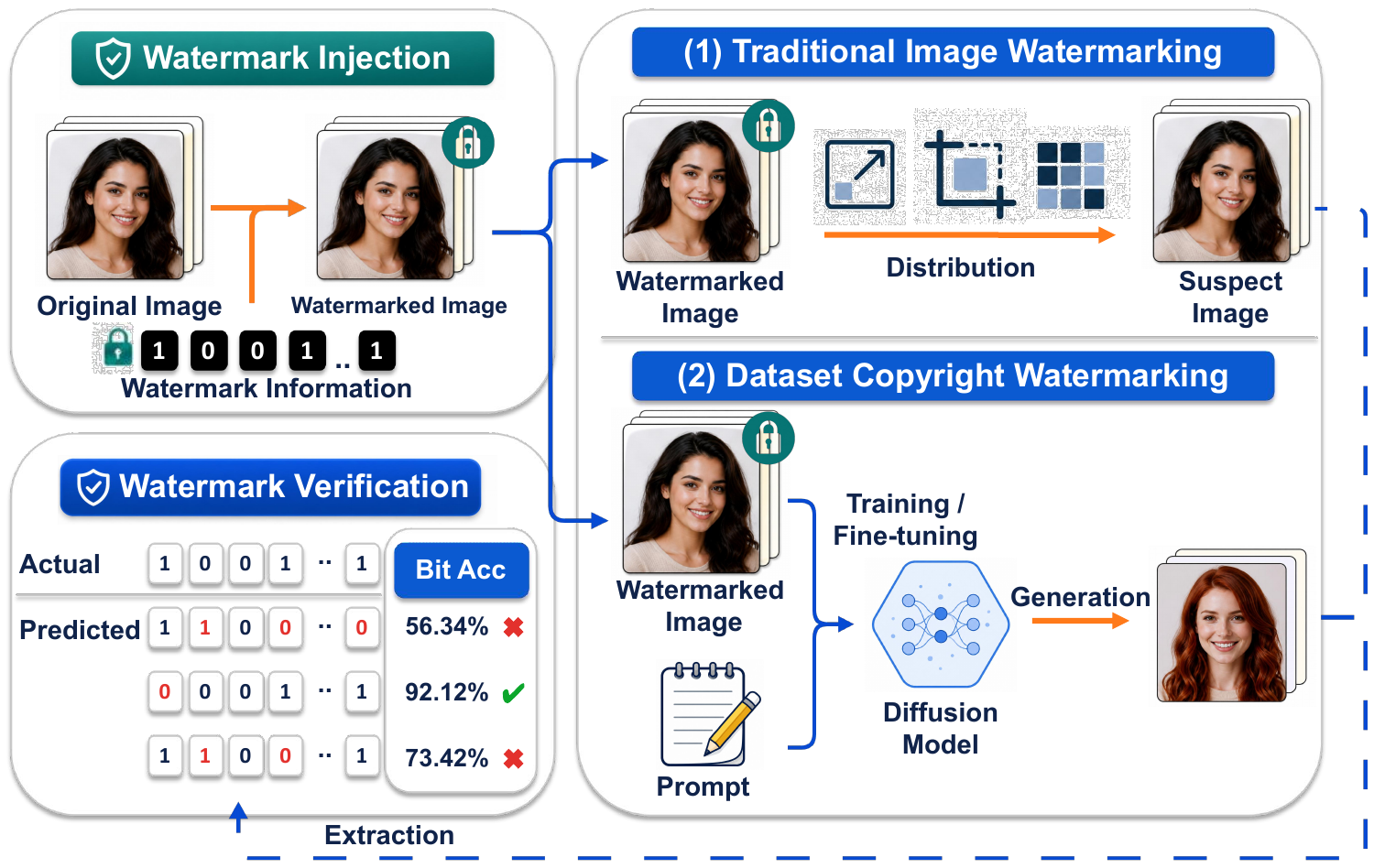}
  \caption{Comparison between the traditional image watermarking and the dataset copyright watermarking. 
  Pipeline (1) represents the traditional image watermarking, whereas pipeline (2) represents the dataset copyright watermarking that we focus on in this paper.}
  \label{fig:framework}
\end{figure}

\mypara{Multi-bit Watermark}
Multi-bit watermarking methods are characterized by their ability to encode and recover a watermark, such as ownership identifiers or metadata, enabling fine-grained and reliable traceability and multi-party license management.

Before the rise of modern generative models, a large body of traditional multi-bit watermarking techniques had already been developed for digital media protection~\cite{ingemar2008digital,petitcolas1999information,begum2020digital}. 
Representative approaches span traditional transform-domain techniques (\eg, \textbf{DwtDct}~\cite{deb2012combined}), which embed bits into wavelet coefficients to trade off imperceptibility against robustness. 
They also encompass learning-based encoder–decoder frameworks, such as \textbf{SSL}~\cite{fernandez2022watermarking} and \textbf{RivaGAN}~\cite{zhang2019robust}, which perform end-to-end optimization for multi-bit embedding and recovery.
However, there exists a significant difference between the traditional image watermarking pipeline and the dataset copyright watermarking pipeline, as illustrated in~\autoref{fig:framework}. 
These methods overlook the indirect ``data-to-model'' transmission channel in dataset copyright auditing, where the watermark must survive training distillation and ultimately manifest in the output distribution. 
Furthermore, experimental results show that these baselines fail to emerge reliably in T2I generations, with bit recovery accuracy often degrading to near random guessing.

Recently, several attempts have been made to tailor multi-bit dataset watermarking for unconditional diffusion-model pipelines, yet significant limitations persist in practical T2I environments. 
\textbf{DiffusionShield}~\cite{cui2023diffusionshield} integrates watermarks by jointly optimizing denoising and reconstruction losses within an unconditional DDPM~\cite{ho2020denoising} framework.
However, it proves fragile under T2I scenarios as dominant semantic alignment gradients inevitably suppress the subtle pixel-level textures required for multi-bit recovery.
Similarly, \textbf{HMark}~\cite{li2025hmark} assigns bits to latent triggers during DDPM fine-tuning. 
However, these static signals are undermined by T2I stochasticity and cross-attention redistribution, resulting in a pronounced trade-off between auditing resolution and image fidelity.
Moreover, the scheme introduced by \textbf{Luo~\etal}~\cite{luo2023steal} leverages bit-recovery statistics as fingerprints for detecting style mimicry. 
However, it does not extend well to scenarios requiring high-capacity message recovery, since it depends on aggregate distributional patterns rather than accurate, bit-level attribution.
Consequently, these approaches remain insufficient for robust, high-capacity dataset auditing under the complex dynamics of black-box T2I generation.

\subsection{Discussion}
\label{subsec:baseline_selection}
In our evaluations, we select DwtDct~\cite{deb2012combined}, SSL~\cite{fernandez2022watermarking}, RivaGAN~\cite{zhang2019robust}, and DiffusionShield~\cite{cui2023diffusionshield} as our primary baselines for the following reasons. 
Although DwtDct~\cite{deb2012combined}, SSL~\cite{fernandez2022watermarking}, and RivaGAN~\cite{zhang2019robust} were originally designed for media distribution, they are frequently repurposed in generative AI contexts as foundational baselines for dataset copyright tracing due to their technical maturity. 
We therefore include them to evaluate whether watermarks embedded as pixel- or frequency-domain perturbations can survive the ``data-to-model'' channel and support downstream copyright tracing.

Regarding diffusion-oriented methods, we choose DiffusionShield~\cite{cui2023diffusionshield} as a representative end-to-end generative baseline.
We exclude HMark~\cite{li2025hmark} from our evaluation because its source code is currently unavailable for reproducible comparison.
Furthermore, while Luo~\etal~\cite{luo2023steal} provide a valuable auditing framework and evaluation perspective, their implemented watermark injection and verification modules still rely on traditional image watermarking methods rather than a distinct end-to-end multi-bit design tailored for high-capacity recovery.  
Therefore, comparing against representative traditional watermarking baselines already covers the core technical components underlying their instantiation.  
For this reason, we do not treat Luo~\etal~\cite{luo2023steal} as a separate, standalone algorithmic baseline.
We also provide a qualitative comparison with HMark~\cite{li2025hmark} and Luo~\etal~\cite{luo2023steal} in~\autoref{sec:baseline_selection}, as their objectives and settings differ from ours.

\section{Problem Definition}

\subsection{Threat Model}
\label{sec:threat_model}
We consider a typical two-party dataset misuse auditing scenario aligned with the existing work~\cite{du2025dataset, huang2024general, shao2025databench, chen2023face}: the \textit{Dataset Owner} (Defender) and the \textit{Model Trainer} (Attacker).
We further elaborate on the respective goals and capabilities of both parties below.

\mypara{Dataset Owner (Defender)}
The owner holds the copyright to the dataset and aims to secure it before public release~\cite{sablayrolles2020radioactive}. 
The core capabilities of the dataset owner are summarized as follows. 
\begin{itemize}
    \item \textbf{Dataset and Provenance Management}: The dataset owner has the ability to preprocess the dataset in order to embed copyright information.
    This can be achieved by inserting designated multi-bit watermarks $m$ into the data samples. 
    \item \textbf{Black-box Auditing}: The dataset owner can conduct post-hoc auditing with only black-box access to suspect models. 
    This involves issuing text prompts, analyzing the synthesized outputs to verify the watermark's presence, and further decoding the embedded identifier information to support both misuse auditing and responsibility tracing.
\end{itemize}

\mypara{Model Trainer (Attacker)}
The trainer, acting as an attacker, seeks to exploit the released dataset to train or fine-tune T2I models while evading copyright attribution~\cite{huang2024general,shao2025databench}. 
Depending on their capability to evade detection, we consider two types of attackers.

\begin{itemize}
\item \textbf{Vanilla Attacker.} 
The vanilla attacker is unaware of which samples are watermarked or which embedding method is used. 
Therefore, the attacker applies generic image-processing operations without exploiting any method-specific knowledge. 
We consider two attack stages.
First, before model training, the attacker may preprocess the protected training images to sanitize the dataset and weaken watermark evidence.
Second, after model deployment, the attacker may post-process the generated outputs before returning them to the auditor, thereby interfering with watermark detection and message recovery~\cite{nie2022diffusion}. 


\item \textbf{Adaptive Attacker.}
We further consider an adaptive attacker who knows the watermarking framework and auditing workflow but lacks the secret message, trained watermark-module parameters, and detector calibration data.
Accordingly, the attacker can design targeted strategies to suppress evidence of watermarking during data preparation and model training.
We evaluate five adaptive attacks: VAE re-encoding~\cite{zhao2023recipe}, image regeneration~\cite{nie2022diffusion}, text re-captioning~\cite{li2022blip}, watermarked-data filtering, and continued fine-tuning~\cite{chen2021refit} on clean data, each aimed at attenuating watermark evidence while retaining sufficient visual and semantic utility for downstream T2I training.
\end{itemize}

\subsection{Dataset Copyright Auditing}
\label{sec:dw_framework}
As shown in the auditing pipeline of~\autoref{fig:framework}, the dataset owner aims to audit whether a dataset has been used to train or fine-tune a suspect model~\cite{shao2025databench}. 
The auditing pipeline consists of two stages: watermark injection and watermark verification.

\mypara{Watermark Injection}
Before the dataset $\mathcal{D}$ is released, the dataset owner injects a secret multi-bit watermark information $m$ into the samples to produce the watermarked dataset $\mathcal{D}^w$.
In the context of T2I models, this injection can occur in the pixel, frequency, or diffusion latent space~\cite{wang2023diagnosis,xie2025training}. 
The primary goal is to ensure that the watermark is learnable by diffusion models while remaining imperceptible, thereby preserving data utility.
Then, an unauthorized model trainer (attacker) collects $\mathcal{D}^w$ and uses it to train or fine-tune a model $M$. 

\mypara{Watermark Verification}
When a model $M$ is suspected of infringement, the dataset owner queries model $M$ with input prompts, collects the generated outputs, and first determines whether the watermark is present in the generated images.
Once the watermark's presence is confirmed, the owner uses a decoder to recover the embedded identifier for tracing.


\section{\sysname}
\label{sec:method_design}
\subsection{Intuition}
The key challenge of multi-bit dataset watermarking for T2I models is not merely to embed decodable information into individual images, but to establish a reliable information channel across dataset protection, downstream model training, and stochastic image generation. 
An imperceptible watermark that is inconsistent with the learned data distribution may be suppressed during training, while watermark evidence retained by the model may still be too weak or unstable to recover from a single generated output~\cite{cui2023diffusionshield}. 
Our key observation is that watermark persistence and recovery should therefore be considered jointly: the protected data should introduce message-consistent evidence that can be absorbed by the downstream model, and the auditing process should consolidate weak but coherent traces exposed across generated samples. 
Guided by this view, \sysname jointly optimizes watermark injection and recovery to balance imperceptibility, learnability, and the reliability of multi-bit attribution.

\begin{figure*}[t]
  \centering
  \includegraphics[width=0.9\textwidth]{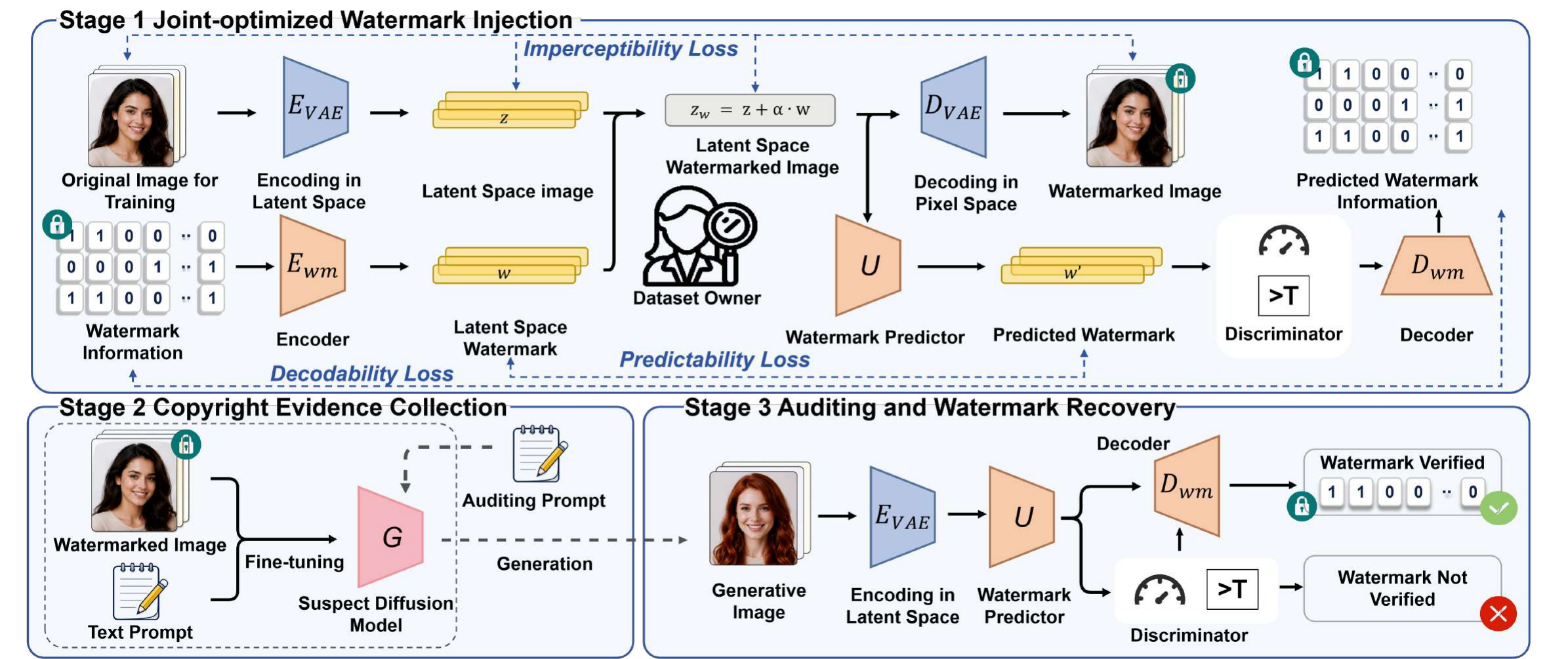}
  \caption{Workflow of \sysname for T2I dataset protection. Stage 1 involves a joint-optimization injection process in which the dataset owner embeds multi-bit watermarks in the latent space, guided by losses for predictability, decodability, and imperceptibility. Stage 2 focuses on collecting copyright evidence by triggering the suspect diffusion model to generate watermarked samples. Stage 3 performs auditing by passing the generated images through the predictor $U$ and the decoder $D_{wm}$ to recover the watermark for verification.}
  \label{fig:workflow}
\end{figure*}

\subsection{Workflow}
\label{sec:method_overview}

The workflow of \sysname is illustrated in~\autoref{fig:workflow}, which contains three stages: (i) joint-optimized watermark injection, (ii) copyright evidence collection, and (iii) auditing and watermark recovery.
Detailed designs for each component are elaborated in~\autoref{sec:design_details}.

\mypara{Stage 1: Joint-optimized Watermark Injection}
In this stage, we train a joint architecture consisting of a latent \textit{encoder-decoder}, a U-Net-based \textit{predictor}, together with a separately trained auditing discriminator for watermark-presence verification.
The encoder learns to generate latent residuals that are recognizable by the predictor, distinguishable by the discriminator, and consistent with the diffusion dynamics.
Meanwhile, the predictor learns to extract watermark residuals from complex latent representations and provide separable evidence for discrimination.
This \textit{pre-alignment} step makes it easier for diffusion models to learn the watermark.
Once optimized, the dataset owner specifies a secret $L$-bit watermark information $m$, which is processed by the trained encoder to produce a latent watermark signal.
These signals are injected into the latent representations of the training samples. 
The resulting watermarked dataset $\mathcal{D}^w$ contains an identity signal tightly coupled to the underlying data distribution.

\mypara{Stage 2: Copyright Evidence Collection}
When a model $M$ is trained on the watermarked dataset $\mathcal{D}^w$, the embedded patterns are manifested in its generated outputs.
Leveraging this property, we collect copyright evidence through black-box queries to the model. 
The dataset owner sends auditing prompts to the model and collects the resulting generated images. 
Given the inherent sampling stochasticity of diffusion models, the watermark signal in a single output may be partially occluded by generative noise. 
To ensure robust verification, we perform multiple queries to obtain a sequence of generated samples. 
This multi-query mechanism allows us to suppress random noise through statistical aggregation, thereby externalizing the model's implicit memory of the training data into detectable signals manifested in the generated images.

\mypara{Stage 3: Auditing and Watermark Recovery}
The final auditing phase performs inference by feeding generated images into a U-Net-based predictor. 
Its multi-scale feature-fusion disentangles subtle watermark signals, which are then verified via a discriminator and hypothesis test before being processed by a decoder for information recovery.
To mitigate sampling stochasticity, we introduce a multi-sample aggregation mechanism.
By applying bit-wise majority voting over independent samples, our method stabilizes recovery to produce a consistent estimate, $\hat{m}$. We further incorporate error correction to enhance identifier recovery and facilitate user-level tracing. 
Finally, comparing this result with the original watermark $m$ enables definitive auditing and attribution decisions.

\subsection{Design Details}
\label{sec:design_details}

This section elaborates on three core technical components of \sysname, which primarily correspond to the two key stages of joint-optimized injection and auditing recovery in~\autoref{sec:method_overview}. Overall, we model the watermark as a structured latent residual aligned with the diffusion process, and organize the design around three aspects: embedding, detection, and aggregated recovery.

\mypara{Latent Space Watermark Embedding}
This component corresponds to the joint-optimized injection stage (Stage 1) in the workflow. Its main goal is to embed the multi-bit watermark into the training data in a learnable, transferable form without disrupting the diffusion generation mechanism.
In latent diffusion models (LDMs), training and sampling occur in the VAE latent space. Watermarks misaligned with this pipeline (\eg, fixed pixel patterns) are often treated as noise and removed during denoising. To ensure representation alignment, we embed the watermark directly into the diffusion-consistent latent $z$, facilitating its propagation into generated content.

Our approach hinges on the joint alignment of embedding location and optimization. Instead of random noise, a learnable encoder generates a content-adaptive structured residual. This encourages the downstream model to absorb the watermark as a learnable latent feature correlated with the data distribution rather than suppressing it as a nuisance.

We employ an adaptive design where the residual is conditioned on both latent content and the $L$-bit message $m$, with a scalar $\alpha$ governing the imperceptibility--decodability trade-off. Specifically, $x$ denotes the original image, $m\in\{0,1\}^{L}$ denotes the binary watermark message, $\mathrm{E}_{\mathrm{VAE}}$ and $\mathrm{D}_{\mathrm{VAE}}$ denote the frozen encoder and decoder of the pretrained VAE, respectively, and $E_{\mathrm{wm}}$ denotes the learnable watermark encoder:
\begin{equation}
\left\{
\begin{aligned}
z_{\phantom{w}} &= \mathrm{E}_{\mathrm{VAE}}(x), \\
z_{w}           &= z + \alpha \cdot E_{\mathrm{wm}}(z,m), \\
x_{w}           &= \mathrm{D}_{\mathrm{VAE}}(z_{w})
\end{aligned}
\right.
\end{equation}
Where $z$ is the latent representation of $x$, $z_w$ is the watermarked latent, $x_w$ is the resulting watermarked image, and $\alpha$ is the embedding-strength coefficient, with larger values indicating stronger watermark perturbations.

\mypara{Watermark Prediction and Detection}
This component connects the joint-optimized injection stage (Stage 1) and the auditing and watermark recovery stage (Stage 3) in the workflow.

During joint optimization, watermarks survive as weak, structured features entangled with image semantics. A U-Net backbone is well-suited for recovery, as its multi-scale architecture and skip connections isolate subtle traces from complex backgrounds. Its role is therefore to recover residual watermark cues from semantically rich content rather than detect visible artifacts.

During auditing and watermark recovery, the dataset owner queries the suspect model and obtains a set of generated images $\{x_i'\}_{i=1}^{N}$, where $N$ is the number of audit queries and $x_i'$ denotes the image obtained from the $i$-th query. 
Each generated image is first mapped to the VAE latent space, and the U-Net predictor extracts watermark-related features. Based on these features, the discriminator produces a continuous watermark-presence score, while the bit decoder produces a candidate $L$-bit message:
\begin{equation}
\begin{cases}
z_i' = E_{\mathrm{VAE}}(x_i'),\\
r_i = U(z_i'),\\
s_i = C(r_i),\\
m^{(i)} = D_{\mathrm{wm}}(r_i),
\end{cases}
\quad i=1,\ldots,N.
\end{equation}
Where $z_i'$ is the latent representation of $x_i'$, $U$ denotes the U-Net watermark predictor, $r_i$ denotes the watermark-related residual feature, $C$ denotes the watermark-presence discriminator, and $D_{\mathrm{wm}}$ denotes the bit decoder. The variable $s_i\in[0,1]$ is the detector score for the $i$-th output, while $m^{(i)}\in\{0,1\}^{L}$ is its decoded candidate message.
The final auditing decision is made by performing a hypothesis test over the score set $\{s_i\}_{i=1}^{N}$.

Crucially, \sysname does not require access to or knowledge of the image-to-latent module, including the VAE architecture or parameters, used by the downstream generation model.
Because the watermark is a content-adaptive structured residual, it leaves traces in the image domain that remain recoverable even after re-encoding by different VAEs.

\mypara{Multi-sample Aggregation}
This component corresponds to the auditing recovery stage (Stage 3), transforming evidence from multiple black-box queries into a stable consensus result. Given that diffusion sampling is inherently stochastic~\cite{karras2022elucidating,song2020denoising}, decoded bits from a single sample may fluctuate even if the suspect model has successfully captured the watermark. To enhance stability and reproducibility, we query the model multiple times and decode an $L$-bit message from each output. By aggregating these predictions bit-wise, our approach effectively suppresses decoding errors induced by sampling randomness, ensuring a more robust recovery.

Let $m^{(i)}$ denote the predicted $L$-bit message extracted from the $i$-th generated image, for $i=1,\dots,N$, where $N$ is the audit sample size (\ie, the number of queries). We apply an aggregation operator $\textsf{Agg}(\cdot)$ to obtain the estimate $\hat{m}$~\cite{dietterich2000ensemble}:
\begin{equation}
\hat{m} = \textsf{Agg}\left(\{m^{(i)}\}_{i=1}^{N}\right).
\end{equation}
Where $\textsf{Agg}(\cdot)$ denotes the bit-wise multi-sample aggregation operator and $\hat{m}$ denotes the aggregated message estimate.
In our default instantiation, $\textsf{Agg}(\cdot)$ performs bit-wise majority voting to derive a statistically robust consensus from the stochastic query sequence:
\begin{equation}
\hat{m}_j = \mathbb{I}\!\left[\sum_{i=1}^{N}\mathbb{I}\!\left(m^{(i)}_j=1\right) \ge \left\lceil \frac{N}{2}\right\rceil \right], \quad j=1,\dots,L.
\end{equation}
Where $m_j^{(i)}$ and $\hat{m}_j$ denote the $j$-th bits of the individual and aggregated messages, respectively; and $\mathbb{I}[\cdot]$ is the indicator function.
This strategy reduces the impact of occasional single-sample bit errors, yielding a more stable recovered message for auditing.

\subsection{Joint Optimization Strategy}
As shown in Stage~1 of~\autoref{fig:workflow}, we optimize three loss functions to achieve the training goals of predictability, decodability, and imperceptibility.

\mypara{Predictability Loss}
This loss enforces the predictability objective, \ie, the U-Net predictor $U(\cdot)$ should directly recover the embedded watermark signal itself from the watermarked latent.
We treat the encoder-produced signal $E_{\mathrm{wm}}(z,m)$ as the target and require the predictor output on the watermarked latent to match it.
This regression objective forces $U(\cdot)$ to disentangle the structured watermark perturbation from complex latent content, yielding a stable intermediate representation for subsequent bit decoding.
Concretely, we first construct the watermarked latent and then minimize the squared $\ell_2$ error between the predicted watermark signal and the ground-truth watermark signal:
\begin{equation}
\begin{aligned}
z_w &= z + \alpha \cdot E_{\mathrm{wm}}(z,m) \\
\mathcal{L}_{\text{pre}} &= \left\lVert U(z_w) - E_{\mathrm{wm}}(z,m) \right\rVert_2^2~,
\end{aligned}
\end{equation}
where $\alpha$ controls the embedding strength. 
For a fixed $\alpha$, $\mathcal{L}_{\text{pre}}$ directly measures how well the predictor can recover the injected perturbation.

\mypara{Decodability Loss}
This loss enforces the decodability objective: even if the predictor can approximate the watermark signal pattern, the recovered signal must be information-bearing so that the watermark information $m$ can be reliably decoded.
We thus supervise the bit decoder $\mathrm{D}_{\mathrm{wm}}(\cdot)$ to recover the $L$-bit message from the predicted signal $U(z_w)$.
We adopt a per-bit binary cross-entropy loss:
\begin{equation}
\mathcal{L}_{\text{bit}}=\mathrm{BCE}\big(\mathrm{D}_{\mathrm{wm}}({U(z_w)}),\, m\big).
\end{equation}
$\mathcal{L}_{\text{pre}}$ and $\mathcal{L}_{\text{bit}}$ are complementary: the former ensures the watermark signal is recoverable, while the latter ensures that the recovered signal supports faithful message decoding.

\mypara{Imperceptibility Loss}
To preserve the imperceptibility of the model generation, we jointly constrain the watermark perturbation in both the latent and pixel spaces.
First, since diffusion training and denoising dynamics operate in latent space, excessive deviation from the original latent $z$ may distort the latent distribution and degrade generation quality.
We thus regularize the latent modification magnitude by penalizing $\|z_w - z\|_2^2$.
Second, to directly suppress visible deviations in the image domain, we decode the watermarked latent with the VAE decoder and constrain pixel-level deviation from the original image.
Formally, we define:
\begin{equation}
\begin{aligned}
\mathcal{L}_{\text{imp}}
&= \|z_w - z\|_2^2
+ \big\|\mathrm{D}_{\mathrm{VAE}}(z_w) - x\big\|_2^2, \\
z_w &= z + \alpha \cdot E_{\mathrm{wm}}(z,m).
\end{aligned}
\end{equation}
This combined loss controls perturbation energy in the latent space while constraining watermark imperceptibility, thereby improving stealthiness without sacrificing predictability or decodability.

\mypara{Two-stage Training Schedule}
A central challenge in training \sysname is the conflict between channel establishment and imperceptibility control, especially at high watermark capacities.
The losses $\mathcal{L}_{\text{pre}}$ and $\mathcal{L}_{\text{bit}}$ encourage the model to form a decodable watermark channel, whereas $\mathcal{L}_{\text{imp}}$ penalizes visible and latent perturbations.
If all three objectives are optimized jointly from the start, the easiest way to reduce $\mathcal{L}_{\text{imp}}$ is often to suppress the embedding signal before a reliable channel has been established, leading to a degenerate solution that favors imperceptibility over predictability and decodability.

To resolve this conflict, we adopt a two-stage schedule.
Instead of optimizing all objectives equally from the outset, we first solve a simpler subproblem: establishing a watermark channel that is both predictable and decodable.
Only after this channel becomes stable do we introduce imperceptibility regularization to refine the solution.
This can be viewed as a staged or continuation optimization strategy: stage 1 finds a feasible decoding solution, and stage 2 optimizes the full objective around it.

During warm-up, we optimize only the verifiability-related losses, \ie, $\mathcal{L}_{\text{pre}}$ and $\mathcal{L}_{\text{bit}}$, so that the encoder-predictor-decoder pathway converges to a regime where the watermark can be predicted and the $L$-bit message can be decoded.
After warm-up, we enable $\mathcal{L}_{\text{imp}}$ and optimize the full objective:
\begin{equation}
\left\{
\begin{aligned}
\mathcal{L}_{\text{train}}
&=
\lambda_{1}\mathcal{L}_{\text{pre}}
+
\lambda_{2}\mathcal{L}_{\text{bit}},
&& \text{if } \text{epoch} < \tau(t), \\
\mathcal{L}_{\text{train}}
&=
\lambda_{1}\mathcal{L}_{\text{pre}}
+
\lambda_{2}\mathcal{L}_{\text{bit}}
+
\lambda_{3}\mathcal{L}_{\text{imp}},
&& \text{otherwise}.
\end{aligned}
\right.
\label{eq:two_stage_schedule}
\end{equation}

Empirically, we do not observe destabilization after the switch; instead, warm-up reduces early collapse and is particularly beneficial in high-capacity settings.
We define $\tau(t)$ as the warm-up cutoff, implemented either as fixed epochs (default) or adaptively based on $\mathcal{L}_{\text{bit}}$ convergence.
In practice, a brief warm-up is sufficient to stabilize message recovery before full-objective training improves imperceptibility while preserving verification performance.

\subsection{Technical Novelty of Our Proposal}
To address the challenges of watermark persistence, semantic interference, and stochastic generation, \sysname combines three complementary designs: structured latent alignment to preserve watermark evidence, generative denoising to separate watermark residuals from visual semantics, and multi-query aggregation with BCH error correction for reliable provenance recovery.

\mypara{Structured Latent Alignment for Watermark Persistence}
Instead of introducing a pixel-domain pattern that may be treated as independent noise, \sysname represents the provenance message as a message-conditioned and content-adaptive residual in the latent space used by the diffusion model.
Therefore, although the watermark component is attenuated according to the diffusion schedule, it remains a structured, diffusion-scaled displacement rather than immediately becoming independent Gaussian noise.
Since samples carrying the same provenance message contain correlated message-dependent residuals, they can contribute a coherent component to the downstream denoising objective instead of behaving as unrelated random perturbations~\cite{gumemorization}. 
This makes the watermark more likely to be learned together with the protected data distribution and subsequently manifested in generated outputs.

\mypara{Semantic Disentanglement through Generative Denoising} 
The challenge is to extract watermarks from generated images whose prompts and semantics differ from the protected training data. 
In such outputs, the watermark appears as a weak residual entangled with visual content, making direct bit decoding vulnerable to semantic interference.
To address this issue, \sysname introduces a U-Net-based predictor supervised by the embedded residual.
Rather than directly predicting bits, it reconstructs watermark-related evidence from the generated image latent. 
We formulate recovery as a denoising task that suppresses semantic variation and isolates the structured residual before decoding. 
The U-Net's multi-scale features and skip connections facilitate this separation~\cite{ho2020denoising}, reducing dependence on prompt-specific semantics.

\mypara{Reliable Provenance Recovery from Stochastic Outputs}
Diffusion sampling introduces substantial variability across generated outputs.
Even when a downstream model retains the watermark, a single generated image may expose only partial evidence, causing bit predictions to fluctuate across queries. 
\sysname therefore queries the suspect model multiple times and aggregates complementary evidence from independently generated images through bit-wise majority voting. 
This aggregation suppresses sample-specific decoding errors and produces a more stable consensus message.
We further apply BCH error correction to correct the remaining bit errors and map the recovered codeword to a provenance identifier. 
Together, these three designs enable \sysname to transform weak, stochastic evidence from individual outputs into reliable identifier-level provenance.


\section{Implementation}
\label{sec:impl}

\mypara{Datasets and Models}
We evaluate \sysname on three datasets spanning distinct visual domains and three widely used T2I backbones.
For datasets, we use (i) \textbf{Pok\'{e}mon}~\cite{pinkney2022pokemon}, which exhibits a consistent cartoon-style distribution; (ii) \textbf{CelebA}~\cite{liu2015deep}, paired with facial captions generated by LLaVA; and (iii) \textbf{WikiArt}~\cite{saleh2015large}, which contains diverse artistic styles and complex classical textures.
For model backbones, we consider (i) \textbf{Stable Diffusion v1.5} (SD v1.5)~\cite{rombach2022high} as a standard latent diffusion baseline, (ii) \textbf{Stable Diffusion XL} (SDXL)~\cite{podell2023sdxl} with a higher-capacity architecture and dual text encoders, and (iii) \textbf{Kandinsky 2.2}~\cite{razzhigaev2023kandinsky}, an image-prior-based diffusion model.
Following black-box auditing, we fine-tune backbones on watermarked datasets using LoRA~\cite{hu2021lora}.
Additional dataset statistics and implementation details are deferred to~\autoref{sec:appendix_setup}.

\mypara{Reference VAE}
For joint optimization, \sysname uses the pretrained AutoencoderKL from SD v1.5 as a fixed reference VAE. 
Its encoder and decoder remain frozen throughout training, and their parameters are neither retrained nor fine-tuned. 
The reference VAE need not exactly match the image-to-latent module of the downstream T2I model. 
SDXL also employs an AutoencoderKL, but it uses different pretrained weights and a latent scaling factor of 0.13025, compared with 0.18215 for SD v1.5, whereas Kandinsky 2.2 employs a MoVQ encoder-decoder. 
We empirically validate the generalization of the reference VAE across different backbones, including SD v1.5, SDXL, and Kandinsky 2.2 in~\autoref{subsec:generalization}.

\mypara{Prompts and Generations}
For each dataset, we construct a pool of $150$ unique prompts.
These prompts are generated by a Large Language Model~\cite{li2023synthetic} while maintaining a linguistic style similar to that of the original datasets, ensuring they are semantically representative of the target domains.

To simulate a post hoc auditing scenario, we query the suspect model with black-box access. 
In this setting, the auditor is restricted to obtaining images generated by the model in response to text prompts. 
We then decode the watermarks from these output images to verify whether the training data has been used without authorization. 
To obtain a robust estimate of the audit success rate, we employ a statistical sampling protocol in which $N=30$ prompts are randomly sampled from the pool in each audit trial to query the model.
This procedure is repeated $100$ times, and the average success rate is calculated as the final metric to evaluate the reliability of watermark detection.

\mypara{Hyperparameters}
We summarize the hyperparameters used in our experiments. 
We consider message lengths 
$L\in\{8,16,32,64,\allowbreak 128,256,512,1024\}$ bits to study the effect of watermark capacity on decodability and behavior. 
Unless otherwise stated, we use $L=32$ as the message length in most experiments.
The embedding strength $\alpha$ is a scalar that controls the trade-off between imperceptibility and decodability, and we set $\alpha=0.5$ by default.

We optimize a joint objective with weights $(\lambda_1,\lambda_2,\lambda_3)$ and adopt a two-stage training schedule to prevent early degradation in image quality. 
In stage 1, we focus on verifiability-related objectives and set $\lambda_1 = 1$, $\lambda_2 = 1$, and $\lambda_3 = 0$. 
In the later stage 2, we enable the imperceptibility regularizers and set all weights to 1, thereby improving perceptual quality while preserving reliable decoding.

\mypara{Competitors}
We compare \sysname against four representative multi-bit watermarking baselines that cover both diffusion-specific and traditional image watermarking paradigms: DiffusionShield~\cite{cui2023diffusionshield}, DwtDct~\cite{deb2012combined}, SSL~\cite{fernandez2022watermarking}, and RivaGAN~\cite{zhang2019robust}. 
DiffusionShield~\cite{cui2023diffusionshield} is a diffusion-based scheme designed for generative pipelines, and it is the closest baseline to our threat model and deployment setting.
The other three are widely used multi-bit image watermarking methods originally developed for traditional media distribution.
Furthermore, they are frequently employed as baselines for auditing dataset watermarking. 
For a fair comparison, we follow each method's implementation and recommended hyperparameters, and evaluate all baselines under the same datasets, model backbones, and auditing protocol as \sysname.

\mypara{Metrics} 
We evaluate the performance of \sysname from two primary perspectives: \textbf{effectiveness} and \textbf{imperceptibility}.

\begin{itemize}
    \item \textbf{Effectiveness}: This category evaluates whether the watermark can provide reliable evidence for infringement auditing and whether the embedded message can be recovered for tracing.
    \begin{itemize}
        \item \textit{Auditing Decision}: This dimension evaluates whether a suspect model can be identified as having learned the target watermark. For each generated image, the discriminator outputs a watermark-presence score. Given multiple query images in a single audit, we perform statistical analysis on these detector outputs and use a hypothesis test to determine whether the suspect model contains the target watermark. The detailed statistical formulation is provided in~\autoref{subsec:significance}. We report \textbf{TPR} and \textbf{FPR}~\cite{fawcett2006introduction} over fine-tuned watermarked models and clean models trained without the watermark.
        
        \item \textit{Message Recovery}: After watermark presence is confirmed, this dimension evaluates the accuracy and stability of recovering the embedded identifier information from generated outputs. \textbf{Image-level bit accuracy ($\mathrm{BA}_{\text{img}}$)}: We decode the message from each individual image and calculate its bit accuracy relative to the target message $m$, reflecting the per-sample recovery quality. \textbf{Aggregated bit accuracy ($\mathrm{BA}_{\text{agg}}$)}: To simulate practical tracing, we aggregate the predicted bits from multiple queried images into a single consensus message $\hat{m}$ and report its mean accuracy over repeated audit trials.
    \end{itemize}
    
    \item \textbf{Imperceptibility}: This category ensures the watermark injection remains stealthy and does not degrade model utility.
    \begin{itemize}
        \item \textit{Image Fidelity}: Focuses on image-level invisibility. We report \textbf{SSIM}~\cite{wang2004image} and \textbf{LPIPS}~\cite{zhang2018unreasonable} to assess the structural and perceptual fidelity between the original and watermarked samples, ensuring no conspicuous artifacts.
        \item \textit{Generation Quality}: Focuses on model-level performance. We compute \textbf{FID}~\cite{heusel2017gans} to quantify the distribution shift after fine-tuning, ensuring the model preserves its generative capabilities with minimal side effects.
    \end{itemize}
\end{itemize}

\section{Evaluation}

\subsection{Overall Performance}
\mypara{Effectiveness}
Our analysis begins with a comparative evaluation of \sysname against representative multi-bit watermark baselines.
We assess performance on the SD v1.5~\cite{rombach2022high} model across three datasets: Pok\'{e}mon, CelebA, and WikiArt.
Unless otherwise specified, all reported effectiveness results are obtained under a high-fidelity setting where the watermarked images maintain $\mathrm{SSIM}>0.95$, ensuring that auditing does not rely on visually obvious artifacts.
Detailed experimental protocols and imperceptibility analysis are provided in~\autoref{subsec:generalization}.

\textbf{Experiments show that \sysname achieves the best overall effectiveness among all compared methods, consistently combining reliable auditing decisions with stronger message recovery.}
As summarized in~\autoref{tab:effectiveness_sd15_combined}, \sysname attains perfect or near-perfect auditing performance across all three datasets, achieving TPR/FPR of 100.00\%/0.00\% on Pok\'{e}mon and CelebA, and 100.00\%/1.00\% on WikiArt.
Meanwhile, it also yields the best message recovery results, reaching $BA_{img}$/$BA_{agg}$ scores of 85.35\%/100.00\%, 77.52\%/93.88\%, and 80.75\%/95.75\% on Pok\'{e}mon, CelebA, and WikiArt, respectively.
These results indicate that \sysname not only provides reliable evidence for infringement auditing, but also preserves sufficient multi-bit information for downstream tracing.

\textbf{A key advantage of \sysname is that it simultaneously supports both auditing and tracing, whereas existing baselines often succeed in only one of the two aspects.}
For example, DiffusionShield achieves a TPR/FPR of 100.00\%/0.00\% on Pok\'{e}mon, but its message recovery remains essentially random, with $BA_{img}=49.55\%$ and $BA_{agg}=49.43\%$.
Similarly, SSL attains a TPR/FPR of 100.00\%/1.00\% on Pok\'{e}mon, yet its aggregated bit accuracy is only 54.86\%.
These results suggest that some existing methods can provide detectable watermark evidence, but fail to preserve sufficiently accurate embedded messages for reliable identifier recovery.
In contrast, among the evaluated methods, \sysname is the only method that consistently achieves both high auditing reliability and high-fidelity multi-bit recovery across all datasets.


\textbf{Several baseline methods exhibit pathological behavior in message recovery after diffusion fine-tuning, where the decoded bits become unstable or even anti-correlated with the true watermark message.}
Their aggregated bit accuracy frequently stays near random guessing or even falls below 50\%.
For example, DwtDct attains a $BA_{agg}$ of only 30.43\% on Pok\'{e}mon, while RivaGAN yields a $BA_{agg}$ of 44.65\% on CelebA.
This suggests that methods designed primarily for traditional pixel-domain watermarking fail to preserve high-capacity identifiers through the stochastic diffusion training and sampling pipeline.
Consequently, although some of them may still provide partial evidence for watermark presence, they remain inadequate for reliable multi-bit tracing in black-box T2I auditing.

\begin{table}[t]
\centering
\caption{Effectiveness on Stable Diffusion v1.5. Metrics with ($\uparrow$) indicate higher is better; ($\downarrow$) indicate lower is better.}
\label{tab:effectiveness_sd15_combined}
\footnotesize
\setlength{\tabcolsep}{4pt}
\renewcommand{\arraystretch}{0.9}
\begin{tabular}{l|l|cccc}
\toprule
Dataset & Method & TPR (\%)$\uparrow$ & FPR (\%)$\downarrow$ & $BA_{img}$ (\%)$\uparrow$ & $BA_{agg}$ (\%)$\uparrow$ \\
\midrule
\multirow{5}{*}{Pok\'{e}mon}
& DiffusionShield   & \heatPrecFone{100.00}   & \heatFPR{0.00} & \heatBAimg{49.55} & \heatBAimg{49.43} \\
& DwtDct            & \heatPrecFone{2.00}   & \heatFPR{0.00} & \heatBAimg{42.77} & \heatBAimg{30.43} \\
& SSL               & \heatPrecFone{100.00}   & \heatFPR{1.00} & \heatBAimg{55.62} & \heatBAimg{54.86} \\
& RivaGAN           & \heatPrecFone{10.00}   & \heatFPR{1.00} & \heatBAimg{58.54} & \heatBAimg{62.59} \\
& \textbf{\sysname} & \heatPrecFone{100.00} & \heatFPR{0.00} & \heatBAimg{85.35} & \heatBAimg{100.00} \\
\midrule
\multirow{5}{*}{CelebA}
& DiffusionShield   & \heatPrecFone{100.00} & \heatFPR{0.00} & \heatBAimg{49.57} & \heatBAimg{49.43} \\
& DwtDct            & \heatPrecFone{0.00} & \heatFPR{0.00} & \heatBAimg{52.60} & \heatBAimg{59.59} \\
& SSL               & \heatPrecFone{97.00} & \heatFPR{0.00} & \heatBAimg{60.91} & \heatBAimg{67.76} \\
& RivaGAN           & \heatPrecFone{13.00} & \heatFPR{0.00} & \heatBAimg{47.52} & \heatBAimg{44.65} \\
& \textbf{\sysname} & \heatPrecFone{100.00} & \heatFPR{0.00} & \heatBAimg{77.52} & \heatBAimg{93.88} \\
\midrule
\multirow{5}{*}{WikiArt}
& DiffusionShield   & \heatPrecFone{13.00}   & \heatFPR{0.00} & \heatBAimg{49.43} & \heatBAimg{49.43} \\
& DwtDct            & \heatPrecFone{0.00}   & \heatFPR{0.00} & \heatBAimg{51.19} & \heatBAimg{57.47} \\
& SSL               & \heatPrecFone{92.00}   & \heatFPR{4.00} & \heatBAimg{47.89} & \heatBAimg{48.83} \\
& RivaGAN           & \heatPrecFone{100.00}   & \heatFPR{0.00} & \heatBAimg{59.25} & \heatBAimg{71.22} \\
& \textbf{\sysname} & \heatPrecFone{100.00} & \heatFPR{1.00} & \heatBAimg{80.75} & \heatBAimg{95.75} \\
\bottomrule
\end{tabular}
\end{table}

\mypara{Imperceptibility}
We evaluate imperceptibility at two hierarchical levels: data-level image fidelity and model-level generation quality.

\textbf{Experimental results show that \sysname maintains high image fidelity while achieving competitive generation quality, leading to a better overall trade-off when effectiveness is also taken into account.}
As shown in~\autoref{tab:image_quality_combined}, our method consistently preserves high image fidelity across all benchmarks, with SSIM above 0.97 and LPIPS remaining low.
In particular, on the Pok\'{e}mon dataset, \sysname achieves the best generation quality, reaching an FID of 77.23 and significantly outperforming all baselines.
On CelebA and WikiArt, although \sysname does not obtain the lowest FID, it still achieves competitive generation quality while substantially outperforming methods in effectiveness.
In contrast, traditional schemes such as DwtDct~\cite{deb2012combined} and RivaGAN~\cite{zhang2019robust} better preserve visual quality, but their limited message-recovery capability makes them less suitable for reliable auditing and tracing.
Therefore, when effectiveness and imperceptibility are considered jointly, \sysname demonstrates stronger practical utility.

Qualitative results in~\autoref{fig:image} further confirm that both the watermarked images and the outputs of fine-tuned models remain visually indistinguishable from their clean counterparts, without introducing noticeable artifacts or style shifts.
Residual analysis reveals that the watermark signal is not random noise but is highly content-adaptive and tightly coupled to underlying textures.
This intrinsic correlation makes watermark removal difficult without compromising structural integrity, thereby simultaneously ensuring stealthiness and robustness against tampering.
A more detailed analysis is provided in~\autoref{subsec:qualitative_analysis}.

\begin{table}[t]
\centering
\caption{Imperceptibility and generation quality comparison across three datasets. Metrics include SSIM, LPIPS, and FID.}
\label{tab:image_quality_combined}
\footnotesize
\setlength{\tabcolsep}{8pt}
\renewcommand{\arraystretch}{0.9}
\begin{tabular}{l|l|ccc}
\toprule
Dataset & Method & {SSIM( > 0.95) $\uparrow$} & {LPIPS $\downarrow$} & {FID $\downarrow$} \\
\midrule
\multirow{5}{*}{Pok\'{e}mon}
& \cellcolor{mylightblue} DiffusionShield & \cellcolor{mylightblue} 0.9551 & \cellcolor{mylightblue} 0.0392 & \cellcolor{mylightblue} 184.23 \\
& \cellcolor{mylightblue} DwtDct         & \cellcolor{mylightblue} \textbf{0.9993} & \cellcolor{mylightblue} \textbf{0.0002} & \cellcolor{mylightblue} 86.34 \\
& \cellcolor{mylightblue} SSL              & \cellcolor{mylightblue} 0.9536 & \cellcolor{mylightblue} 0.0549 & \cellcolor{mylightblue} 140.04 \\
& \cellcolor{mylightblue} RivaGAN          & \cellcolor{mylightblue} 0.9777 & \cellcolor{mylightblue} 0.0031 & \cellcolor{mylightblue} 86.62 \\
& \cellcolor{mylightblue} \textbf{\sysname} & \cellcolor{mylightblue} 0.9805 & \cellcolor{mylightblue} 0.0112 & \cellcolor{mylightblue} \textbf{77.23} \\
\midrule
\multirow{5}{*}{CelebA}
& \cellcolor{mylightgray} DiffusionShield & \cellcolor{mylightgray} 0.9508 & \cellcolor{mylightgray} 0.0579 & \cellcolor{mylightgray} 235.45 \\
& \cellcolor{mylightgray} DwtDct         & \cellcolor{mylightgray} \textbf{0.9971} & \cellcolor{mylightgray} \textbf{0.0045} & \cellcolor{mylightgray} 87.83 \\
& \cellcolor{mylightgray} SSL              & \cellcolor{mylightgray} 0.9761 & \cellcolor{mylightgray} 0.0574 & \cellcolor{mylightgray} 233.03 \\
& \cellcolor{mylightgray} RivaGAN          & \cellcolor{mylightgray} 0.9852 & \cellcolor{mylightgray} 0.0226 & \cellcolor{mylightgray} \textbf{65.31} \\
& \cellcolor{mylightgray} \textbf{\sysname} & \cellcolor{mylightgray} 0.9790 & \cellcolor{mylightgray} 0.0312 & \cellcolor{mylightgray} 87.14 \\
\midrule
\multirow{5}{*}{WikiArt}
& \cellcolor{mylightblue} DiffusionShield & \cellcolor{mylightblue} 0.9591 & \cellcolor{mylightblue} 0.0265 & \cellcolor{mylightblue} \textbf{93.49} \\
& \cellcolor{mylightblue} DwtDct         & \cellcolor{mylightblue} \textbf{0.9978} & \cellcolor{mylightblue} \textbf{0.0018} & \cellcolor{mylightblue} 106.22 \\
& \cellcolor{mylightblue} SSL              & \cellcolor{mylightblue} 0.9647 & \cellcolor{mylightblue} 0.0311 & \cellcolor{mylightblue} 170.92 \\
& \cellcolor{mylightblue} RivaGAN          & \cellcolor{mylightblue} 0.9872 & \cellcolor{mylightblue} 0.0149 & \cellcolor{mylightblue} 107.41 \\
& \cellcolor{mylightblue} \textbf{\sysname} & \cellcolor{mylightblue} 0.9764 & \cellcolor{mylightblue} 0.0116 & \cellcolor{mylightblue} 109.27 \\
\bottomrule
\end{tabular}
\end{table}

\begin{table}[t]
\centering
\caption{Generalization of \sysname across various datasets and T2I backbones.}
\label{tab:generalization}
\footnotesize 
\setlength{\tabcolsep}{3pt} 
\renewcommand{\arraystretch}{0.9}
\begin{tabular}{l|l|cccc|c}
\toprule
Dataset & Model & TPR (\%)$\uparrow$ & FPR (\%)$\downarrow$ & $BA_{img}$ (\%)$\uparrow$ & $BA_{agg}$ (\%)$\uparrow$ & FID$\downarrow$ \\
\midrule
\multirow{3}{*}{Pok\'{e}mon} & \cellcolor{mylightblue} SD~v1.5   & \cellcolor{mylightblue} 100.00 & \cellcolor{mylightblue} 0.00 & \cellcolor{mylightblue} 85.35 & \cellcolor{mylightblue} 100.00 & \cellcolor{mylightblue} 77.23 \\
& \cellcolor{mylightblue} SDXL      & \cellcolor{mylightblue} 100.00 & \cellcolor{mylightblue} 0.00 & \cellcolor{mylightblue} 78.08 & \cellcolor{mylightblue} 97.50 & \cellcolor{mylightblue} 59.70 \\
& \cellcolor{mylightblue} Kandinsky & \cellcolor{mylightblue} 100.00 & \cellcolor{mylightblue} 2.00 & \cellcolor{mylightblue} 86.12 & \cellcolor{mylightblue} 95.88 & \cellcolor{mylightblue} 105.37 \\\midrule

\multirow{3}{*}{CelebA}
& \cellcolor{mylightgray} SD~v1.5   & \cellcolor{mylightgray} 100.00 & \cellcolor{mylightgray} 0.00 & \cellcolor{mylightgray} 77.52 & \cellcolor{mylightgray} 93.88 & \cellcolor{mylightgray} 87.14 \\
& \cellcolor{mylightgray} SDXL      & \cellcolor{mylightgray} 100.00 & \cellcolor{mylightgray} 0.00 & \cellcolor{mylightgray} 62.33 & \cellcolor{mylightgray} 81.59 & \cellcolor{mylightgray} 81.06 \\
& \cellcolor{mylightgray} Kandinsky & \cellcolor{mylightgray} 100.00 & \cellcolor{mylightgray} 0.00 & \cellcolor{mylightgray} 76.52 & \cellcolor{mylightgray} 85.18 & \cellcolor{mylightgray} 74.34 \\\midrule

\multirow{3}{*}{WikiArt}
& \cellcolor{mylightblue} SD~v1.5   & \cellcolor{mylightblue} 100.00 & \cellcolor{mylightblue} 1.00 & \cellcolor{mylightblue} 80.75 & \cellcolor{mylightblue} 95.75 & \cellcolor{mylightblue} 109.27 \\
& \cellcolor{mylightblue} SDXL      & \cellcolor{mylightblue} 100.00 & \cellcolor{mylightblue} 3.00 & \cellcolor{mylightblue} 76.17 & \cellcolor{mylightblue} 89.09 & \cellcolor{mylightblue} 108.19 \\
& \cellcolor{mylightblue} Kandinsky & \cellcolor{mylightblue} 100.00 & \cellcolor{mylightblue} 0.00 & \cellcolor{mylightblue} 65.29 & \cellcolor{mylightblue} 84.63 & \cellcolor{mylightblue} 104.31 \\
\bottomrule
\end{tabular}
\end{table}
\subsection{Generalization}
\label{subsec:generalization}

Beyond the evaluation on SD v1.5 across three benchmark datasets, we further extend our experiments to more advanced architectures, namely SDXL and Kandinsky 2.2, to assess the generalization capability of \sysname.
This cross-architecture evaluation on the same three datasets is designed to verify whether our auditing framework remains effective across diverse generative backbones.

\mypara{Across Models}
\sysname generalizes well across different T2I backbones.
As shown in~\autoref{tab:generalization}, it achieves a TPR of 100.00\% in all nine dataset--model combinations, while keeping the FPR consistently low, within 0--3\%.
These results indicate that the auditing reliability of \sysname remains highly stable even when the downstream generative backbone changes substantially.

Notably, this cross-architecture success is achieved even when the downstream generative models employ VAE structures different from the $E_{\text{VAE}}$ used during the watermark injection stage.
This suggests that \sysname remains effective under variations in downstream latent representations and does not rely on a specific VAE architecture to maintain auditability.
Although image-level and aggregated bit accuracy vary across backbones, the overall recovery performance remains strong.
For example, on Pok\'{e}mon, \sysname achieves $BA_{\text{agg}}$ values of 100.00\%, 97.50\%, and 95.88\% on SD v1.5, SDXL, and Kandinsky, respectively, demonstrating near-perfect message recovery across architectures.

\mypara{Across Datasets}
Compared with model architecture, data-domain characteristics appear to play a larger role in determining the difficulty of message recovery.
Among the three datasets, Pok\'{e}mon is consistently the easiest, achieving the strongest recovery performance across all backbones.
In contrast, CelebA and WikiArt are more challenging, and their relative difficulty varies across models.

For example, under SDXL, the $BA_{\text{agg}}$ is 97.50\% on Pok\'{e}mon, 81.59\% on CelebA, and 89.09\% on WikiArt.
Under Kandinsky, the corresponding values become 95.88\%, 85.18\%, and 84.63\%.
This suggests that architecture changes do not fundamentally break auditability, but they do affect the fidelity of multi-bit recovery in a dataset-dependent manner.
In particular, domains with more complex fine-grained structures or stronger semantic competition make stable bit preservation more difficult.

\begin{figure*}[htbp] 
    \centering
    \begin{minipage}{0.95\textwidth}
     \begin{subfigure}[b]{0.32\textwidth}
         \centering
         \includegraphics[width=\textwidth]{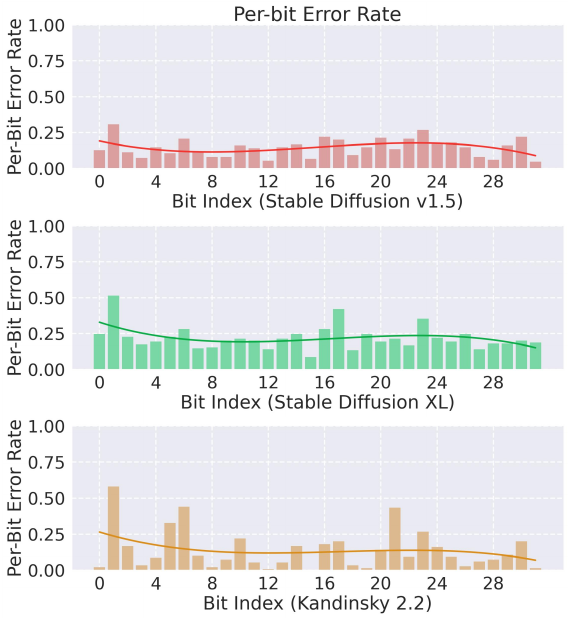}
         \caption{Pok\'{e}mon}
         \label{fig:sub-wm_bit_wise-a}
     \end{subfigure}%
     \hfill
     \begin{subfigure}[b]{0.32\textwidth}
         \centering
         \includegraphics[width=\textwidth]{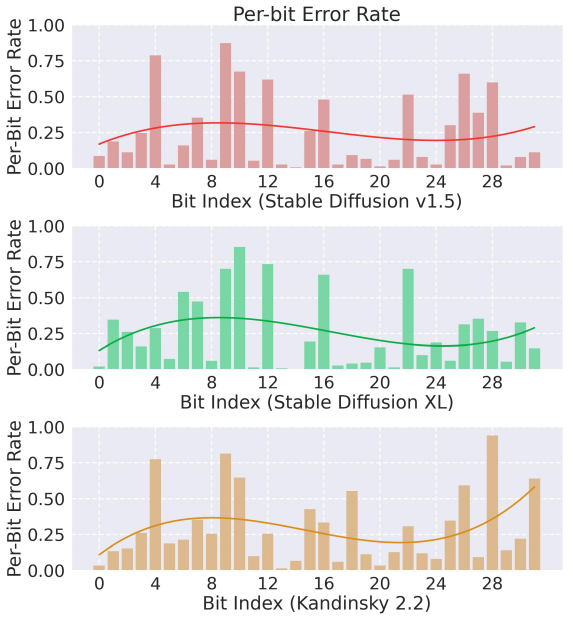}
         \caption{CelebA}
         \label{fig:sub-wm_bit_wise-b}
     \end{subfigure}%
     \hfill
     \begin{subfigure}[b]{0.32\textwidth}
         \centering
         \includegraphics[width=\textwidth]{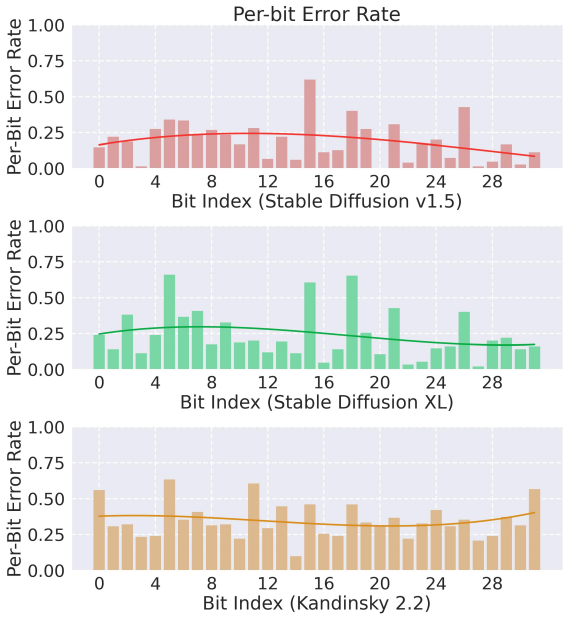}
         \caption{WikiArt}
         \label{fig:sub-wm_bit_wise-c}
     \end{subfigure}
        
     \caption{Average per-bit error rates during watermark extraction across multiple datasets and model backbones.}
     \label{fig:wm_bit_wise}
     \end{minipage}
\end{figure*}

\mypara{Consistency of Error Bits}
The intrinsic characteristics of a data domain induce broadly consistent per-bit error patterns across different model architectures.
As illustrated in~\autoref{fig:wm_bit_wise}, the per-bit error-rate plots show substantial overlap for different backbones on the same dataset.
In particular, several high-error bit positions recur across models, suggesting that certain bits are inherently more vulnerable to corruption due to the interaction between the watermark signal and the data domain.

Although the exact magnitudes vary across architectures, the overall peak locations and fluctuation trends remain broadly consistent.
This cross-architecture consistency indicates that the main bottlenecks in watermark decodability are more likely determined by the data domain than by any specific model architecture.
This observation provides empirical guidance for designing adaptive watermark encoding strategies for challenging datasets.

\subsection{Watermark Configuration}

\mypara{Watermark Rate}
In realistic dataset misuse scenarios, protected samples often account for only a small fraction of the fine-tuning corpus, since model trainers typically aggregate data from multiple sources~\cite{carlini2023extracting}. 
Therefore, a low watermark rate is not an edge case, but an unavoidable practical challenge for dataset copyright auditing.

\begin{figure*}[htbp] 
     \centering
     \begin{minipage}{0.9\textwidth}
         \centering
         \begin{subfigure}[b]{0.32\textwidth}
             \centering
             \includegraphics[width=\textwidth]{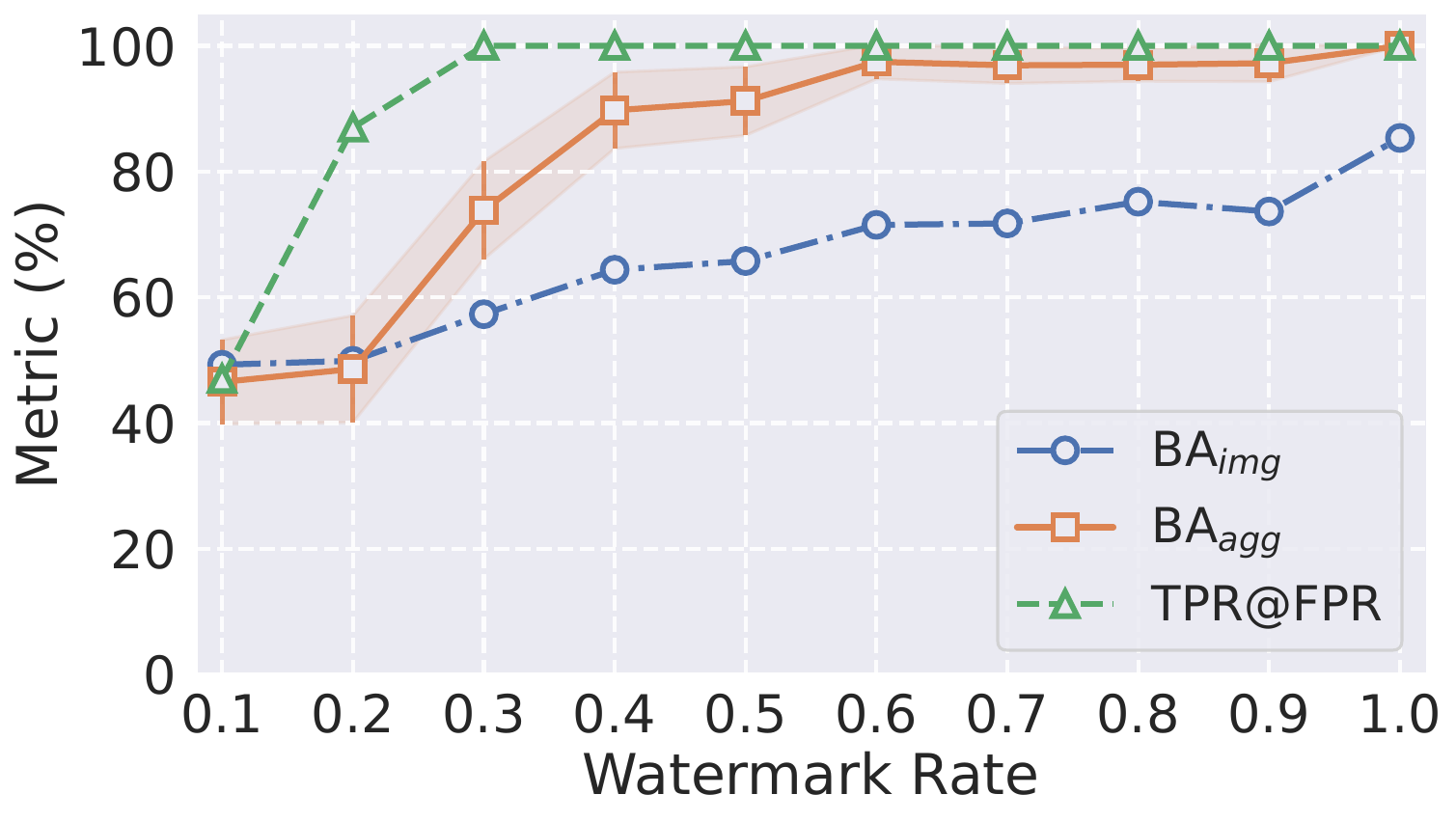}
             \caption{Pok\'{e}mon}
             \label{fig:sub-wm_rate-a}
         \end{subfigure}%
         \hfill
         \begin{subfigure}[b]{0.32\textwidth}
             \centering
             \includegraphics[width=\textwidth]{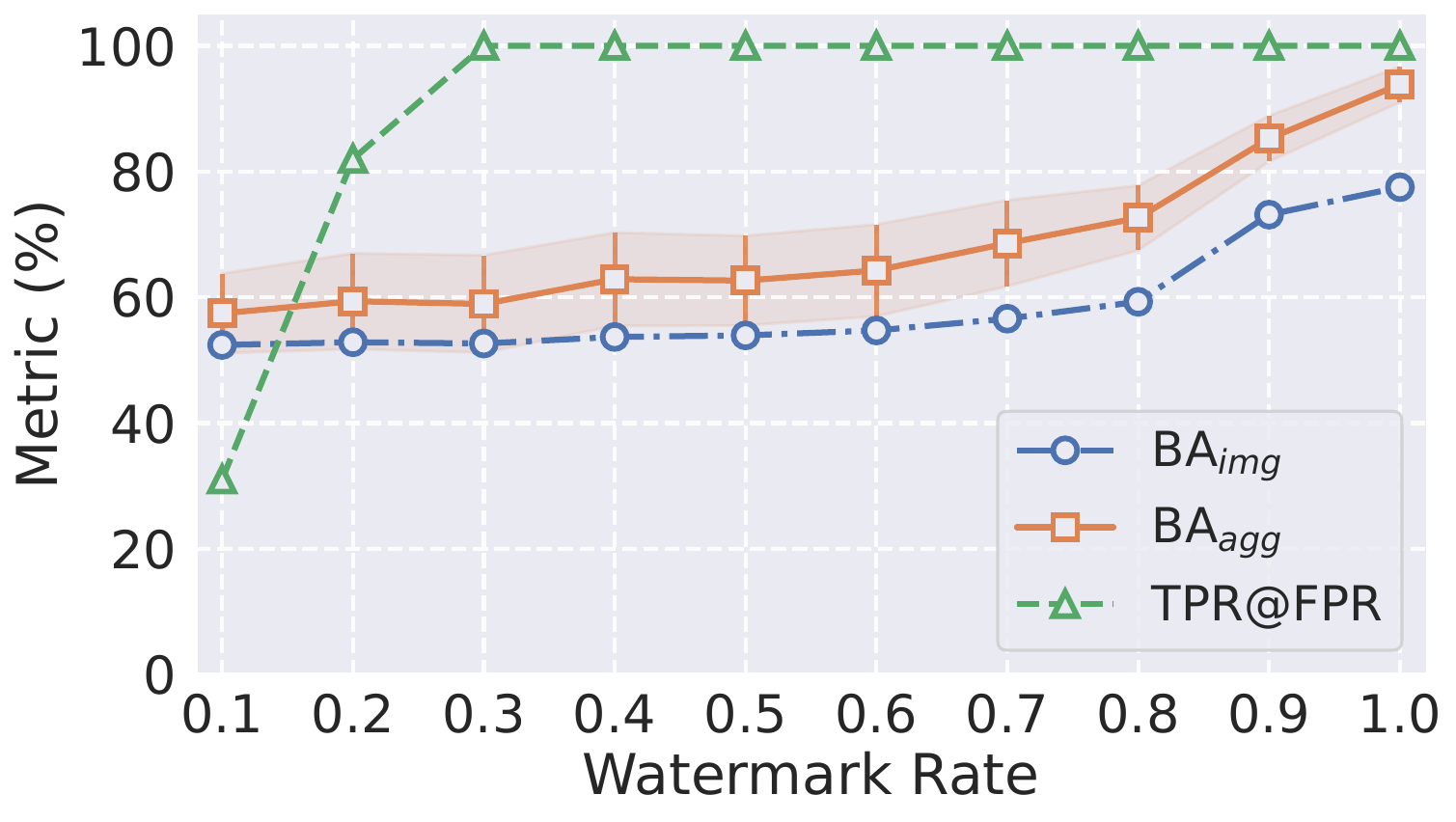}
             \caption{CelebA}
             \label{fig:sub-wm_rate-b}
         \end{subfigure}%
         \hfill
         \begin{subfigure}[b]{0.32\textwidth}
             \centering
             \includegraphics[width=\textwidth]{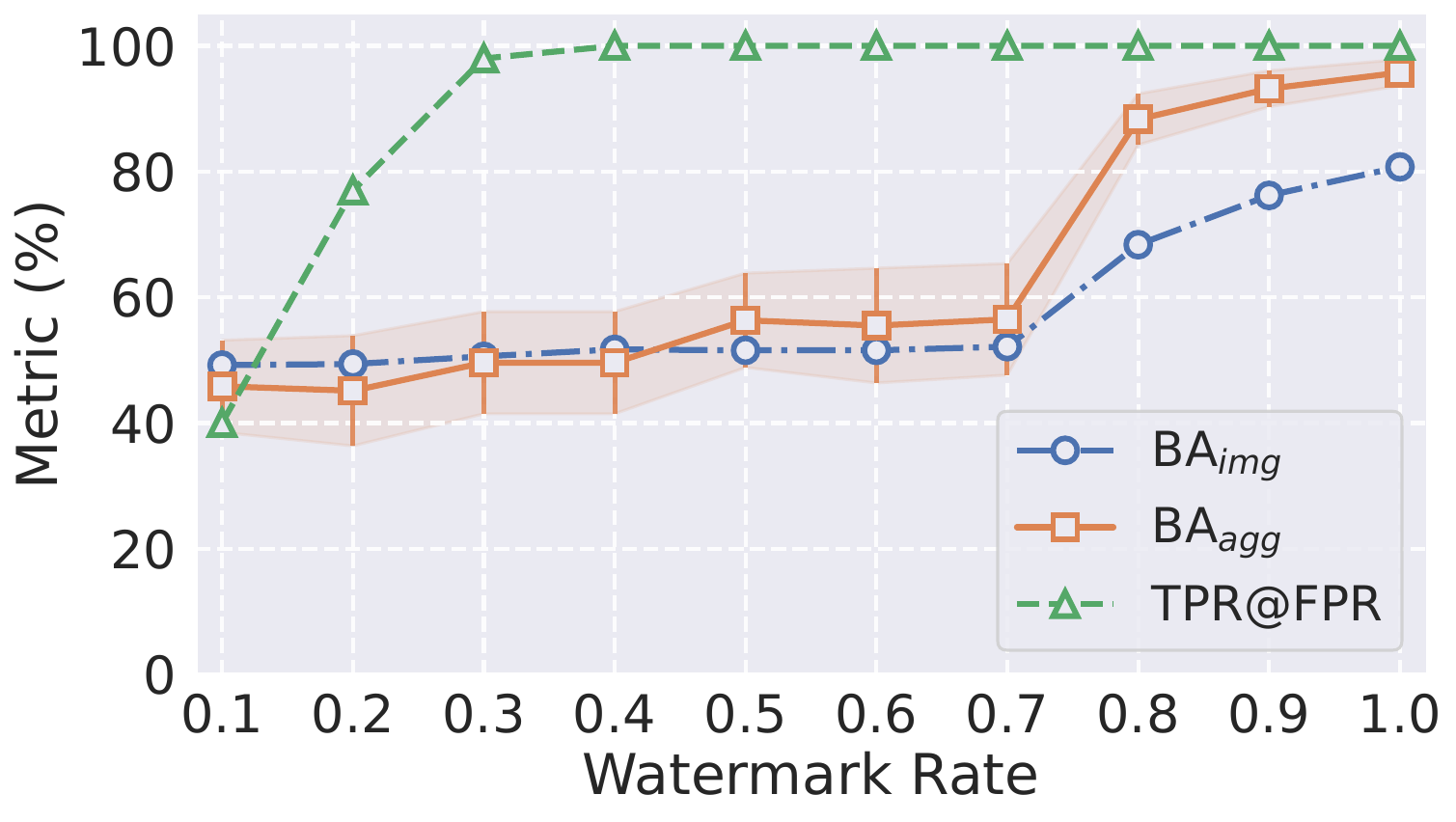}
             \caption{WikiArt}
             \label{fig:sub-wm_rate-c}
         \end{subfigure}
     \end{minipage}

     \caption{Impact of watermark rate $\rho$ on auditing performance. The plots illustrate the evolution of $\mathrm{BA}_{\text{img}}$, $\mathrm{BA}_{\text{agg}}$, and TPR (measured at $\mathrm{FPR}=0$) as the proportion of watermarked training samples increases from 0.1 to 1.0.}
     \label{fig:wm_rate}
\end{figure*}

To examine this setting, we evaluate \sysname under different watermark rates $\rho$ on three datasets, as shown in~\autoref{fig:wm_rate}. 
Overall, lower watermark rates inevitably weaken both auditing evidence and message recovery, because the watermarked samples are diluted by clean data during fine-tuning, thereby reducing their influence on model optimization.

Nevertheless, \sysname still retains meaningful auditing capability at low watermark rates. 
On Pok\'{e}mon and CelebA, the TPR rises from 47.00\% and 31.00\% at $\rho=0.1$ to 87.00\% and 82.00\% at $\rho=0.2$, and reaches 100\% on both datasets from $\rho=0.3$ onward. 
By contrast, WikiArt is more sensitive to the weakening effect of low watermark rates: at low $\rho$, both $\mathrm{BA}_{\text{img}}$ and $\mathrm{BA}_{\text{agg}}$ remain close to random guessing, suggesting that in domains with more complex textures, sparse protected samples are less likely to leave recoverable multi-bit evidence.
A promising future direction is to improve robustness at low watermark rates by designing watermarks that remain effective when protected samples are sparse, such as semantic-aware embedding strategies, more redundancy-efficient encoding schemes, or auditing procedures that accumulate evidence more effectively across queries.

\begin{table}[t]
\centering
\caption{Impact of watermark strength $\alpha$ on effectiveness and imperceptibility trade-offs on the Pok\'{e}mon dataset.}
\label{tab:wm_strength}
\footnotesize
\setlength{\tabcolsep}{4.0pt}
\renewcommand{\arraystretch}{0.9}
\begin{tabular}{c|c c c c|c c c}
\toprule
$\alpha$ & TPR (\%)$\uparrow$ & FPR (\%)$\downarrow$ & $BA_{img}$ (\%)$\uparrow$ & $BA_{agg}$ (\%)$\uparrow$ & SSIM$\uparrow$ & LPIPS$\downarrow$ & FID$\downarrow$ \\
\midrule
0.1 & \heatPrecFone{6.00}    & \heatFPR{0.00} & \heatBAimgn{49.40} & \heatBAimgn{45.19} & \heatSSIM{0.9828} & \heatLPIPS{0.0102} & \heatFID{79.48} \\
0.2 & \heatPrecFone{100.00}  & \heatFPR{0.00} & \heatBAimgn{60.69} & \heatBAimgn{80.88} & \heatSSIM{0.9824} & \heatLPIPS{0.0104} & \heatFID{80.59} \\
0.3 & \heatPrecFone{100.00}  & \heatFPR{0.00} & \heatBAimgn{73.71} & \heatBAimgn{97.84} & \heatSSIM{0.9819} & \heatLPIPS{0.0106} & \heatFID{74.47} \\
0.4 & \heatPrecFone{100.00}  & \heatFPR{0.00} & \heatBAimgn{82.13} & \heatBAimgn{98.09} & \heatSSIM{0.9813} & \heatLPIPS{0.0108} & \heatFID{77.88} \\
0.5 & \heatPrecFone{100.00}  & \heatFPR{0.00} & \heatBAimgn{85.35} & \heatBAimgn{100.00} & \heatSSIM{0.9805} & \heatLPIPS{0.0112} & \heatFID{77.23} \\
0.6 & \heatPrecFone{100.00}  & \heatFPR{0.00} & \heatBAimgn{82.60} & \heatBAimgn{98.91} & \heatSSIM{0.9795} & \heatLPIPS{0.0117} & \heatFID{79.87} \\
0.7 & \heatPrecFone{100.00}  & \heatFPR{0.00} & \heatBAimgn{87.96} & \heatBAimgn{100.00} & \heatSSIM{0.9785} & \heatLPIPS{0.0122} & \heatFID{81.25} \\
0.8 & \heatPrecFone{100.00}  & \heatFPR{0.00} & \heatBAimgn{88.92} & \heatBAimgn{100.00} & \heatSSIM{0.9773} & \heatLPIPS{0.0130} & \heatFID{83.46} \\
0.9 & \heatPrecFone{100.00}  & \heatFPR{0.00} & \heatBAimgn{87.27} & \heatBAimgn{99.91} & \heatSSIM{0.9760} & \heatLPIPS{0.0138} & \heatFID{86.81} \\
1.0 & \heatPrecFone{100.00}  & \heatFPR{0.00} & \heatBAimgn{90.54} & \heatBAimgn{100.00} & \heatSSIM{0.9745} & \heatLPIPS{0.0148} & \heatFID{95.81} \\
\bottomrule
\end{tabular}
\end{table}

\mypara{Watermark Strength}
We investigate watermark strength $\alpha$, which controls the latent watermark signal and governs the trade-off between watermark effectiveness and imperceptibility. 
\autoref{tab:wm_strength} summarizes auditing performance, image fidelity, and generation quality under different $\alpha$ values.

At low strength ($\alpha=0.1$), the watermark is too weak to overcome latent-space noise. 
In this case, the image-level bit accuracy $\mathrm{BA}_{\text{img}}$ drops to 49.40\% and the aggregated bit accuracy $\mathrm{BA}_{\text{agg}}$ remains at 45.19\%, yielding a TPR of only 6.00\%. 
Once $\alpha$ reaches 0.2, the auditing decision becomes reliable, with TPR/FPR reaching 100\%/0\%, although message recovery is still limited ($\mathrm{BA}_{\text{agg}}=80.88\%$).

As $\alpha$ further increases, message recovery improves substantially. 
For example, $\mathrm{BA}_{\text{agg}}$ rises to 97.84\% at $\alpha=0.3$. 
At the same time, increasing $\alpha$ from 0.1 to 1.0 steadily decreases SSIM and increases LPIPS, indicating progressively stronger visual distortion, while overly large $\alpha$ also leads to worse FID. 
In particular, FID becomes noticeably worse when $\alpha$ approaches 1.0.

Therefore, the best trade-off is achieved when $\alpha$ lies in the range of 0.3 to 0.5. 
In this regime, \sysname consistently achieves perfect auditing performance (TPR/FPR = 100\%/0\%), while maintaining aggregated bit accuracy between 97.84\% and 100.00\% and preserving strong image fidelity and competitive generation quality.
To further examine the joint effects of watermark rate and watermark strength, we evaluate \sysname under different combinations of $\rho$ and $\alpha$ on all three datasets.
The complete results are reported in~\autoref{subsec:joint_rate_strength}.

\begin{figure}[htbp] 
     \centering
     \begin{subfigure}[b]{0.49\columnwidth}
         \centering
         \includegraphics[width=\columnwidth]{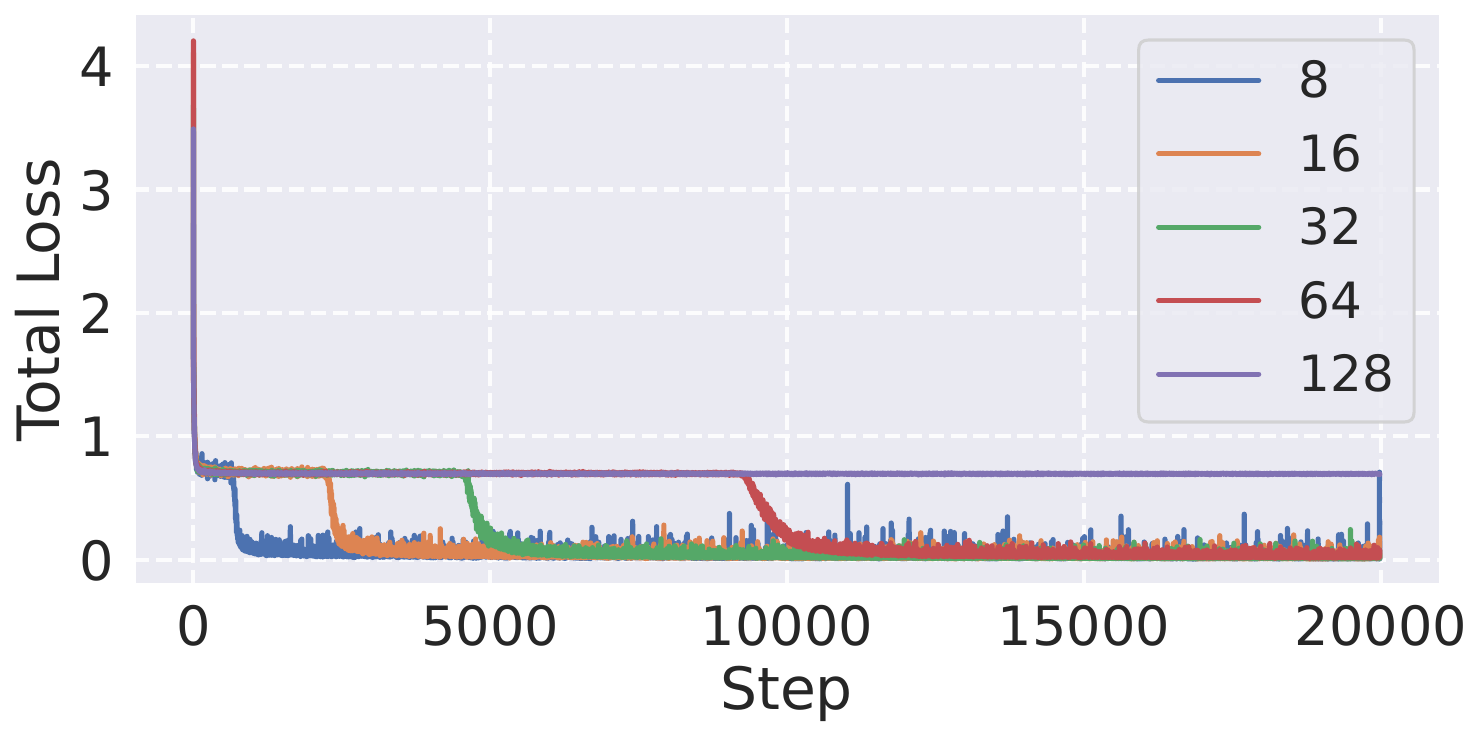}
         \caption{Without two-stage design}
         \label{fig:loss-baseline}
     \end{subfigure}%
     \hfill
     \begin{subfigure}[b]{0.49\columnwidth}
         \centering
         \includegraphics[width=\columnwidth]{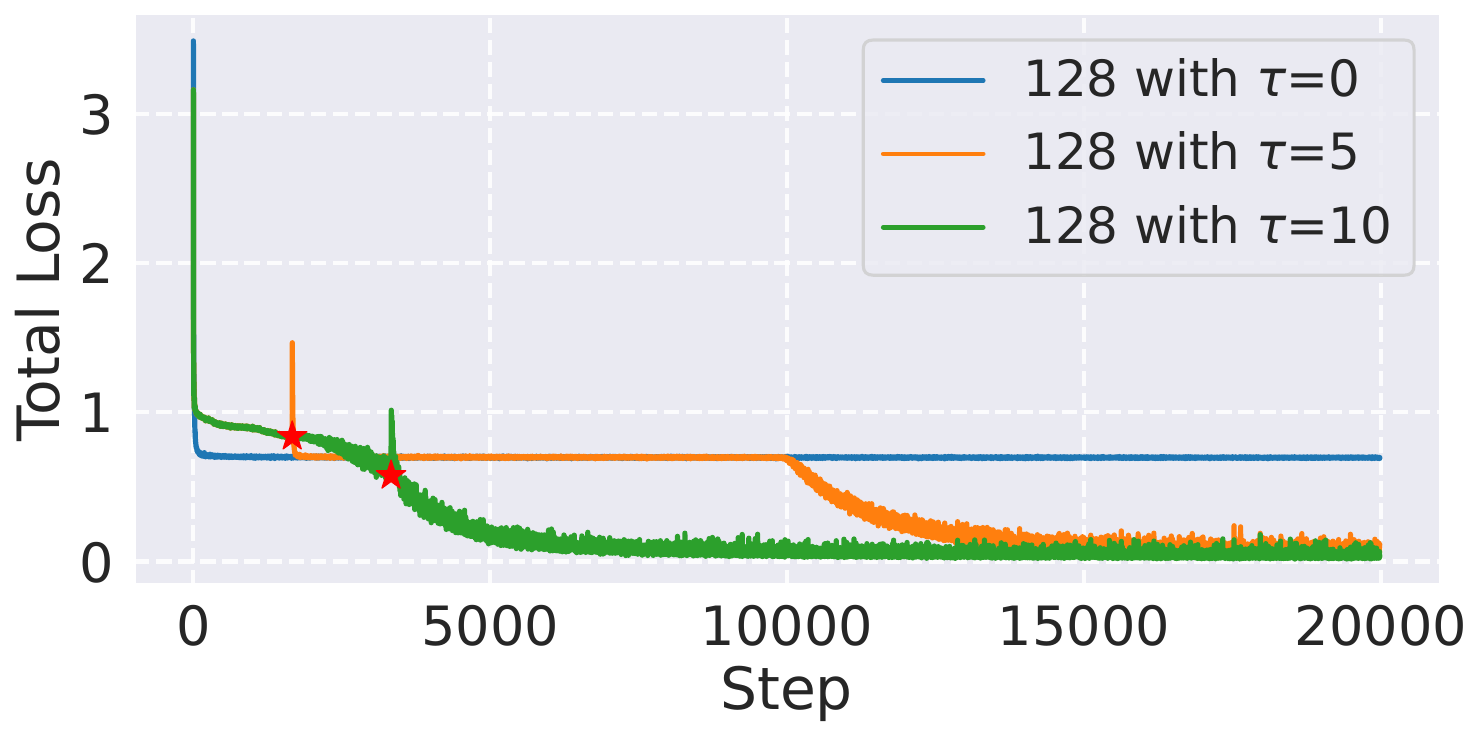}
         \caption{With two-stage design}
         \label{fig:loss-proposed}
     \end{subfigure}%
     \caption{Evolution of watermark extraction loss for varying watermark lengths $L$ under different training configurations on the Pok\'{e}mon dataset.}
     \label{fig:loss}
\end{figure}

\begin{table}[t]
\centering
\caption{Impact of watermark length $L$ of \sysname\ on the Pok\'{e}mon dataset.}
\label{tab:wm_capacity}
\footnotesize
\setlength{\tabcolsep}{6.0pt}
\renewcommand{\arraystretch}{0.9}
\begin{tabular}{l|cccc|c}
\toprule
Bits ($L$) & TPR (\%)$\uparrow$ & FPR (\%)$\downarrow$ & $BA_{img}$ (\%)$\uparrow$ & $BA_{agg}$ (\%)$\uparrow$ & FID$\downarrow$ \\
\midrule
8    & \heatPrecFone{100.00} & \heatFPR{3.00} & \heatBAimgn{89.50} & \heatBAimgn{100.00} & \heatFID {87.39} \\
16   & \heatPrecFone{100.00} & \heatFPR{0.00} & \heatBAimgn{86.08} & \heatBAimgn{100.00} & \heatFID {90.89} \\
32   & \heatPrecFone{100.00} & \heatFPR{0.00} & \heatBAimgn{85.35} & \heatBAimgn{100.00} & \heatFID {77.23} \\
64   & \heatPrecFone{100.00} & \heatFPR{1.00} & \heatBAimgn{79.45} & \heatBAimgn{97.22} & \heatFID {81.51} \\
128  & \heatPrecFone{100.00} & \heatFPR{1.00} & \heatBAimgn{76.46} & \heatBAimgn{95.30} & \heatFID {91.77} \\
256  & \heatPrecFone{100.00} & \heatFPR{0.00} & \heatBAimgn{74.83} & \heatBAimgn{95.43} & \heatFID {87.31} \\
512  & \heatPrecFone{100.00} & \heatFPR{0.00} & \heatBAimgn{73.34} & \heatBAimgn{93.46} & \heatFID {92.31} \\
1024 & \heatPrecFone{100.00} & \heatFPR{2.00} & \heatBAimgn{67.04} & \heatBAimgn{77.72} & \heatFID {100.71} \\
\bottomrule
\end{tabular}
\end{table}

\subsection{High-Capacity Watermarking and Tracing}
\mypara{Watermark Capacity}
Watermark capacity is a key metric for auditing systems, as it determines how much metadata or identifier information can be embedded~\cite{zhang2019steganogan}. We evaluate scalability by varying the watermark length $L$ from 8 to 1024 bits. To support high-capacity recovery, we adopt a two-stage training design that decouples watermark embedding from residual learning.

We first validate the necessity of this design. 
As shown in~\autoref{fig:loss-baseline}, without the two-stage schedule ($\tau=0$), the extraction loss fails to converge once $L>128$, making the watermark unrecoverable. 
In contrast,~\autoref{fig:loss-proposed} shows that the proposed schedule yields lower losses and more stable convergence, while a larger $\tau$ leads to faster convergence. 
This confirms that the two-stage design alleviates optimization difficulty in the high-capacity regime.

Based on this design, we further evaluate performance across different capacities in~\autoref{tab:wm_capacity}.
As $L$ increases from 8 to 512 bits, $\mathrm{BA}_{\text{img}}$ declines from 89.50\% to 73.34\%, reflecting the increasing difficulty of long-message recovery under diffusion stochasticity. 
Nevertheless, at $L=512$, \sysname still achieves $\mathrm{BA}_{\text{agg}}=93.46\%$ with TPR$=100.00\%$ and FPR$=0.00\%$. 
Even at the extreme capacity of 1024 bits, the system maintains $\mathrm{BA}_{\text{agg}}=77.72\%$, indicating that substantial watermark evidence remains in the generated outputs. 
Overall, \sysname provides information injection and recovery across diverse capacity settings, demonstrating its effectiveness for high-capacity auditing.

\mypara{Error Correction for Tracing}
While high aggregated bit accuracy indicates strong watermark recoverability, it does not by itself guarantee identifier-level attribution.
In fine-grained auditing, the recovered bit sequence must be further mapped to a reliable user-level message, such as a licensee ID or user fingerprint.
To support this end-to-end attribution goal, we incorporate an error-correction mechanism on top of the recovered multi-bit watermark.

For each watermark length $L$, we sample a random $K$-bit payload as a user identifier and encode it into an $L$-bit BCH codeword.
As shown in~\autoref{tab:error_correction}, the ECC-enhanced pipeline achieves highly reliable end-to-end tracing across varied watermark lengths.
The tracing success rate (TSR) reaches 100.00\% for $L\in\{8,16,32,64,128\}$, remains 99.00\% at $L=256$, and is still 86.00\% at $L=512$.
These results indicate that although raw bit recovery becomes harder as the watermark length grows, the recovered codewords remain sufficiently close to valid ECC codewords, allowing the decoder to correct residual errors and recover the embedded user identity.

It is worth noting that the reported $BA_{agg}$ values after introducing ECC may be slightly lower than those in the raw watermark-recovery setting.
This does not imply weaker tracing capability.
Rather, the embedded messages are no longer arbitrary random $L$-bit strings, but structured ECC codewords generated from random $K$-bit payloads, and all redundancy bits are also counted in bit-accuracy evaluation.
As a result, small residual errors may slightly reduce raw bit accuracy while still remaining fully correctable by the ECC decoder.
Overall, combining \sysname with ECC bridges watermark recovery and practical attribution, converting high-confidence auditing evidence into exact user-level tracing.
Notably, ECC-based tracing is not evaluated at the extreme capacity of $L=1024$, since the corresponding $BA_{agg}$ is only around 70\%, which is insufficient for reliable ECC decoding.

\begin{table}[t]
\centering
\caption{End-to-end tracing performance with ECC under different watermark lengths. 
$(L, K, D)$ denotes the codeword length, payload length, and minimum Hamming distance, respectively. 
$BA_{agg}$ is measured before ECC decoding, and TSR reports the tracing success rate after ECC decoding.}
\label{tab:error_correction}
\footnotesize
\setlength{\tabcolsep}{4pt}
\renewcommand{\arraystretch}{0.9}
\begin{tabular}{c|c|ccc|c}
\toprule
\multirow{2}{*}{Bit ($L$)} & \multirow{2}{*}{ECC $(L,K,D)$}
& \multicolumn{3}{c|}{$BA_{agg}(\%)\uparrow$}
& \multirow{2}{*}{TSR (\%)$\uparrow$} \\
\cmidrule(lr){3-5}
& & User0 & User1 & User2 & \\
\midrule
8
& $(8,4,3)$
& ${\heatBAimgn{100.00}}_{\scriptscriptstyle \pm 0.00}$
& ${\heatBAimgn{100.00}}_{\scriptscriptstyle \pm 0.00}$
& ${\heatBAimgn{100.00}}_{\scriptscriptstyle \pm 0.00}$
& \heatPrecFone{100.00} \\

16
& $(16,11,3)$
& ${\heatBAimgn{100.00}}_{\scriptscriptstyle \pm 0.00}$
& ${\heatBAimgn{100.00}}_{\scriptscriptstyle \pm 0.00}$
& ${\heatBAimgn{100.00}}_{\scriptscriptstyle \pm 0.00}$
& \heatPrecFone{100.00} \\

32
& $(32,8,9)$
& ${\heatBAimgn{98.44}}_{\scriptscriptstyle \pm 2.32}$
& ${\heatBAimgn{99.50}}_{\scriptscriptstyle \pm 1.23}$
& ${\heatBAimgn{97.25}}_{\scriptscriptstyle \pm 2.39}$
& \heatPrecFone{100.00} \\

64
& $(64,22,13)$
& ${\heatBAimgn{95.67}}_{\scriptscriptstyle \pm 2.21}$
& ${\heatBAimgn{95.92}}_{\scriptscriptstyle \pm 1.81}$
& ${\heatBAimgn{98.80}}_{\scriptscriptstyle \pm 1.58}$
& \heatPrecFone{100.00} \\

128
& $(128,28,27)$
& ${\heatBAimgn{95.48}}_{\scriptscriptstyle \pm 1.55}$
& ${\heatBAimgn{96.96}}_{\scriptscriptstyle \pm 1.40}$
& ${\heatBAimgn{94.94}}_{\scriptscriptstyle \pm 1.22}$
& \heatPrecFone{100.00} \\

256
& $(256,67,45)$
& ${\heatBAimgn{93.36}}_{\scriptscriptstyle \pm 1.68}$
& ${\heatBAimgn{95.20}}_{\scriptscriptstyle \pm 1.52}$
& ${\heatBAimgn{95.49}}_{\scriptscriptstyle \pm 1.47}$
& \heatPrecFone{99.00} \\

512
& $(512,77,95)$
& ${\heatBAimgn{92.38}}_{\scriptscriptstyle \pm 1.62}$
& ${\heatBAimgn{92.03}}_{\scriptscriptstyle \pm 1.21}$
& ${\heatBAimgn{92.12}}_{\scriptscriptstyle \pm 1.16}$
& \heatPrecFone{86.00} \\

1024
& -
& ${\heatBAimgn{78.52}}_{\scriptscriptstyle \pm 0.62}$
& ${\heatBAimgn{76.08}}_{\scriptscriptstyle \pm 1.09}$
& ${\heatBAimgn{69.65}}_{\scriptscriptstyle \pm 1.91}$
& \cellcolor{MorandiRed}{-} \\
\bottomrule
\end{tabular}
\end{table}


\begin{DIFnomarkup}
\begin{table}[t]
\centering
\caption{Robustness of \sysname\ under vanilla  attacks, \ie, pre-processing and output post-processing image perturbations.
JPEG, Res., GBlur, Hue, and Sharp. denote JPEG compression, image resizing, Gaussian blur, hue adjustment, and sharpness adjustment, respectively.
The attack parameters are JPEG quality $=50$, resizing scale $=0.5$, Gaussian blur radius $=2.0$, hue shift $\Delta h=0.1$, and sharpness factor $=2.0$.}

\label{tab:robustness_pre_post}
\footnotesize
\setlength{\tabcolsep}{5.2pt}
\renewcommand{\arraystretch}{0.9}
\begin{tabular}{l|cccc|c}
\toprule
Attack
& TPR (\%)$\uparrow$ & FPR (\%)$\downarrow$ & $BA_{img}$ (\%)$\uparrow$ & $BA_{agg}$ (\%)$\uparrow$ & FID$\downarrow$ \\
\midrule
No Attack & \heatPrecFone{100.00}& \heatFPR{0.00} & \heatBAimgn{85.35} & \heatBAimgn{100.00} & \heatFID{77.23} \\
\midrule
\multicolumn{6}{c}{\textbf{Pre-processing Perturbations}} \\
\midrule 
JPEG & \heatPrecFone{100.00} & \heatFPR{0.00} & \heatBAimgn{75.25} & \heatBAimgn{95.84} & \heatFID{91.38} \\
Resizing & \heatPrecFone{100.00} & \heatFPR{0.00} & \heatBAimgn{79.56} & \heatBAimgn{97.50} & \heatFID{85.01} \\
Blur & \heatPrecFone{83.00} & \heatFPR{0.00} & \heatBAimgn{67.04} & \heatBAimgn{85.41} & \heatFID{135.91} \\
Hue & \heatPrecFone{100.00} & \heatFPR{0.00} & \heatBAimgn{79.33} & \heatBAimgn{99.88} & \heatFID{79.53} \\
Sharpness & \heatPrecFone{100.00} & \heatFPR{0.00} & \heatBAimgn{85.04} & \heatBAimgn{99.72} & \heatFID{87.98} \\
\midrule
\multicolumn{6}{c}{\textbf{Post-processing Perturbations}} \\
\midrule
JPEG & \heatPrecFone{100.00} & \heatFPR{0.00} & \heatBAimgn{71.54} & \heatBAimgn{98.18} & \heatFID{97.70} \\
Resizing & \heatPrecFone{100.00} & \heatFPR{0.00} & \heatBAimgn{85.00} & \heatBAimgn{100.00} & \heatFID{109.36} \\
Blur & \heatPrecFone{100.00} & \heatFPR{0.00} & \heatBAimgn{82.58} & \heatBAimgn{99.94} & \heatFID{140.63} \\
Hue & \heatPrecFone{100.00} & \heatFPR{0.00} & \heatBAimgn{84.06} & \heatBAimgn{99.84} & \heatFID{93.06} \\
Sharpness & \heatPrecFone{100.00} & \heatFPR{0.00} & \heatBAimgn{84.32} & \heatBAimgn{99.97} & \heatFID{94.12} \\
\bottomrule
\end{tabular}
\end{table}
\end{DIFnomarkup}

\begin{DIFnomarkup}
\begin{table}[t]
\centering
\caption{
Robustness of \sysname\ under five adaptive attacks.
For image regeneration, SD 1.5 and SD 2.1 denote the regeneration models. 
The higher regeneration intensity $s$ and latent noise level $l$ correspond to stronger attacks. 
$d_0$ and $d_1$ denote the filtering classifiers trained using the Anime Chibi dataset and a pristine Pok\'emon subset as auxiliary data, respectively.
Re-captioning and Fine-tuning denote text re-captioning and continued fine-tuning, respectively. }

\label{tab:adaptive_robustness}
\footnotesize
\setlength{\tabcolsep}{1.5pt}
\renewcommand{\arraystretch}{1.0}
\begin{tabular}{l c|cccc|c}
\toprule
Attack
& Configuration
& TPR (\%)$\uparrow$
& FPR (\%)$\downarrow$
& $BA_{img}$ (\%)$\uparrow$
& $BA_{agg}$ (\%)$\uparrow$
& FID$\downarrow$ \\
\midrule
\multirow{6}{*}{\shortstack[l]{Image\\Regeneration}}
& SD 2.1, $s = 0.2$
& \heatPrecFone{100.00}
& \heatFPR{0.00}
& \heatBAimgn{67.44}
& \heatBAimgn{90.78}
& \heatFID{86.60} \\

& SD 2.1, $s = 0.3$
& \heatPrecFone{100.00}
& \heatFPR{0.00}
& \heatBAimgn{62.58}
& \heatBAimgn{81.25}
& \heatFID{87.88} \\

& SD 2.1, $s = 0.4$ 
& \heatPrecFone{95.00}
& \heatFPR{0.00}
& \heatBAimgn{59.02}
& \heatBAimgn{74.53}
& \heatFID{90.50} \\

\cmidrule(lr){2-7}

& SD 1.5, $s = 0.2$
& \heatPrecFone{100.00}
& \heatFPR{0.00}
& \heatBAimgn{67.94}
& \heatBAimgn{90.28}
& \heatFID{89.57} \\

& SD 1.5, $s = 0.3$
& \heatPrecFone{98.00}
& \heatFPR{0.00}
& \heatBAimgn{61.29}
& \heatBAimgn{78.59}
& \heatFID{91.11} \\

& SD 1.5, $s = 0.4$
& \heatPrecFone{92.00}
& \heatFPR{4.00}
& \heatBAimgn{58.92}
& \heatBAimgn{74.12}
& \heatFID{92.35} \\

\midrule
\multirow{2}{*}{\shortstack[l]{VAE\\Re-encoding}}
& $l = 0.0$
& \heatPrecFone{100.00}
& \heatFPR{0.00}
& \heatBAimgn{84.50}
& \heatBAimgn{99.87}
& \heatFID{77.12} \\

& $l = 0.15$
& \heatPrecFone{100.00}
& \heatFPR{0.00}
& \heatBAimgn{80.52}
& \heatBAimgn{98.72}
& \heatFID{100.03} \\
\midrule
Re-captioning
& --
& \heatPrecFone{100.00}
& \heatFPR{0.00}
& \heatBAimgn{84.71}
& \heatBAimgn{99.35}
& \heatFID{95.54} \\
\midrule
\multirow{2}{*}{\shortstack[l]{Watermarked\\Data Filter}}
& $d_0$
& \heatPrecFone{100.00}
& \heatFPR{0.00}
& \heatBAimgn{85.29}
& \heatBAimgn{100.00}
& \heatFID{80.53} \\

& $d_1$
& \heatPrecFone{100.00}
& \heatFPR{0.00}
& \heatBAimgn{84.43}
& \heatBAimgn{99.84}
& \heatFID{78.42} \\
\midrule
Fine-tuning
& --
& \heatPrecFone{32.00}
& \heatFPR{2.00}
& \heatBAimgn{50.98}
& \heatBAimgn{57.37}
& \heatFID{155.72} \\

\bottomrule
\end{tabular}
\end{table}
\end{DIFnomarkup}
\section{Robustness}
To evaluate the reliability of \sysname in real-world scenarios, we categorize our robustness evaluation into two classes based on the capabilities of the model trainer defined in the threat model in~\autoref{sec:threat_model}: \textbf{Vanilla Evasion Attacks} and \textbf{Adaptive Evasion Attacks}.
The former covers standard image-level perturbations to test whether watermark evidence can survive common data manipulations.
The latter models an informed attacker employing targeted strategies to evade auditing~\cite{kutter1999fair,voloshynovskiy2001attack,an2024waves}.

\subsection{Vanilla Evasion Attacks}
We evaluate the robustness of \sysname under both pre-processing and output post-processing attacks, with each setting covering five image perturbations: JPEG compression, resizing, Gaussian blur, hue adjustment, and sharpness adjustment.
Pre-processing attacks sanitize protected training images before fine-tuning, whereas post-processing attacks transform generated images before watermark verification.
\autoref{tab:robustness_pre_post} summarizes the resulting auditing performance on the Pok\'{e}mon dataset using Stable Diffusion v1.5.

Under pre-processing attacks, \sysname remains robust against most common perturbations.
For JPEG compression, resizing, hue adjustment, and sharpness adjustment, it maintains TPR$=100.00\%$ and FPR$=0.00\%$.
Although these perturbations reduce $\mathrm{BA}_{\text{img}}$ to varying degrees, $\mathrm{BA}_{\text{agg}}$ remains between 95.84\% and 99.88\%, demonstrating the effectiveness of multi-sample aggregation.
Their FID values remain relatively moderate, ranging from 79.53 to 91.38.
In contrast, Gaussian blur is the most harmful pre-processing attack: it reduces $\mathrm{BA}_{\text{img}}$ and $\mathrm{BA}_{\text{agg}}$ to 67.04\% and 85.41\%, respectively, and lowers the TPR to 83.00\%.
Blur also increases FID to 135.91, indicating that suppressing watermark evidence comes at a substantial cost to data utility.

Under post-processing attacks, \sysname exhibits stronger robustness.
All five perturbations retain TPR$=100.00\%$ and FPR$=0.00\%$, while $\mathrm{BA}_{\text{agg}}$ remains between 98.18\% and 100.00\%.
Even JPEG compression, which produces the lowest post-processing $\mathrm{BA}_{\text{img}}$ of 71.54\%, still achieves an aggregated bit accuracy of 98.18\%.
Gaussian blur also preserves a $\mathrm{BA}_{\text{agg}}$ of 99.94\%, although it causes the largest quality degradation with an FID of 140.63.
These results indicate that once the watermark has been learned and propagated to generated outputs, common image transformations cannot effectively prevent auditing without noticeably degrading output quality.
Overall, \sysname is more sensitive to perturbations applied before model training than to those applied after image generation.


\subsection{Adaptive Evasion Attacks}

To evaluate the resilience of \sysname against strategic purification attempts, we investigate its performance under five sophisticated adversarial scenarios: VAE re-encoding, image regeneration, text re-captioning, watermarked-data filtering, and continued fine-tuning.
The experimental results are summarized in~\autoref{tab:adaptive_robustness}.

\mypara{VAE Re-encoding}
We investigate the VAE re-encoding attack~\cite{zhao2023recipe}, where the attacker perturbs the latent representation with random noise $\epsilon$ during re-encoding to conceal watermark evidence.
Despite this manipulation, \sysname exhibits minor degradation in watermark recoverability: when the perturbation intensity $l$ increases from 0 to 0.15, $BA_{agg}$ decreases only slightly from 99.87\% to 98.72\%. 
In contrast, FID rises sharply from 77.12 to 100.03. 
This indicates that latent-space perturbation fails to remove the watermark unless the attacker is willing to incur substantial perceptual distortion.

\mypara{Image Regeneration}
We further evaluate an image regeneration attack~\cite{nie2022diffusion}, in which an attacker attempts to purify watermarked images by passing them through a secondary diffusion-based reconstruction process.
As the regeneration strength $s$ increases, we observe a monotonic decline in watermark recoverability together with a degradation in generation quality. On SD v2.1, $BA_{agg}$ decreases from 90.78\% ($s=0.2$) to 81.25\% ($s=0.3$) and 74.53\% ($s=0.4$), while FID increases from 86.60 to 87.88 and 90.50.
These results show that although regeneration can weaken the recovered watermark bits, doing so requires increasingly aggressive reconstruction that visibly harms image fidelity, thereby limiting its practical attractiveness as a purification attack.

\mypara{Text Re-captioning}
We evaluate a text re-captioning attack in which the attacker uses BLIP~\cite{li2022blip} to replace the original captions of protected training images before LoRA fine-tuning.
This attack attempts to weaken the association between the watermark evidence and the original image-text pairs by modifying the semantic supervision used during training.
Nevertheless, \sysname remains consistently highly robust, achieving a TPR of 100.00\% with an FPR of 0.00\%.
The resulting $\mathrm{BA}_{\text{img}}$ and $\mathrm{BA}_{\text{agg}}$ remain at 84.71\% and 99.35\%, respectively, while the FID is 95.54.
These results indicate that the watermark is primarily carried by the protected visual content and can still be reliably learned when the associated textual descriptions are rewritten.

\mypara{Watermarked Data Filter}
We evaluate a specialized adaptive scenario in which the model trainer is aware of the underlying watermarking algorithm but lacks access to the training parameters or the specific information.
The attacker first trains a surrogate watermark encoder using auxiliary data and subsequently develops a binary classifier based on the ResNet-34~\cite{he2016deep} architecture to distinguish protected samples.
To construct the training set for this classifier, we use 200 auxiliary images, comprising 100 watermarked samples (generated by the surrogate encoder) and 100 pristine samples, and we discard the detected images.
For auxiliary datasets, we utilize: (i) an Anime Chibi dataset~\cite{phimsiri_anime_chibi_kaggle} with visual styles analogous to the target Pok\'{e}mon data, and (ii) a pristine Pok\'{e}mon subset.

The classifier trained on Anime Chibi detects no protected samples, whereas the classifier trained on the Pok\'{e}mon subset identifies only 23 of the 633 watermarked images.
After removing the detected samples and fine-tuning on the filtered datasets, \sysname\ still achieves TPR/FPR of $100.00\%/0.00\%$ under both settings.
The corresponding $BA_{agg}$ values remain $100.00\%$ and $99.84\%$, with FID values of $80.53$ and $78.42$, respectively.
These results indicate that, without the training parameters or embedded information, a method-aware filter cannot reliably identify enough protected samples to suppress downstream watermark auditing.

\mypara{Continued Fine-tuning}
We further consider a continued fine-tuning attack, where the attacker trains the watermarked model for an additional 20 epochs on the clean Anime Chibi dataset in an attempt to overwrite the learned watermark evidence.
Among the evaluated adaptive attacks, this strategy causes the largest reduction in watermark recoverability: $\mathrm{BA}_{\text{img}}$ and $\mathrm{BA}_{\text{agg}}$ decrease to 50.98\% and 57.37\%, respectively, while the TPR falls to 32.00\% with an FPR of 2.00\%.
However, the FID between outputs generated before and after continued fine-tuning reaches 155.72, indicating a substantial shift from the original model’s generation distribution and therefore a considerable loss of model utility.

\section{Related Work}
\mypara{Data-Use Auditing}
Beyond dataset watermarking, several para\-digms have been developed to audit whether and how particular data are used during model training.
Canary-based auditing~\cite{carlini2019secret,nasr2023tight} inserts a small number of specially designed samples or triggers and examines whether their influence can be detected in the trained model.
Dataset-inference~\cite{maini2021dataset} and membership-inference methods~\cite{shokri2017membership,carlini2022membership} provide statistical evidence about whether a target dataset or individual samples participated in training, while data-attribution methods~\cite{koh2017understanding,park2023trak} estimate the contribution of individual training samples to model behavior.
These paradigms are complementary to dataset watermarking and may require modifying fewer samples, or no samples at all.
However, they generally return binary decisions, statistical evidence, or attribution scores rather than an explicit provenance message.
Therefore, our evaluation focuses on multi-bit watermarking methods that share the objective of message recovery.

\mypara{Diffusion-Aware Image Watermarking}
Recent studies have explored image watermarking in the context of diffusion models, either by leveraging diffusion processes for watermark embedding or by improving robustness against diffusion-based removal. 
ZoDiac~\cite{zhang2024attack} is a zero-bit image watermarking that optimizes an image-specific Stable Diffusion latent containing a predefined Fourier-domain watermark pattern.
It verifies watermark presence through DDIM inversion and statistical testing rather than recovering an explicit message.
LaWa~\cite{rezaei2024lawa} integrates latent-space watermark embedding into the image-generation process, while DERO~\cite{fang2024dero} improves the robustness of directly watermarked images against diffusion-model erasure. 
These methods verify the watermark directly from the protected image or its attacked version after diffusion-based removal or other image transformations. 
In contrast, \sysname embeds an identifier into a training dataset and recovers it from images generated by a separately fine-tuned T2I model, thereby enabling explicit provenance-message transmission and recovery through the indirect ``data-to-model-to-image'' channel.

\section{Limitations}
Despite the encouraging results across different datasets, model architectures, watermark capacities, and attack settings, \sysname still has several limitations that merit further investigation:

(1) \textbf{Evaluation Scope and Transferability}: Our evaluation primarily focuses on LoRA due to its ubiquity in downstream T2I customization. 
Although \sysname does not rely on LoRA-specific mechanisms, broader fine-tuning paradigms remain to be further evaluated. 
Moreover, training dataset-specific encoder-decoder modules incurs non-trivial computational costs, motivating future work on meta-learning or universal encoders.

(2) \textbf{Expansion of Defensive Boundaries}: Although \sysname is resilient in most scenarios, its recoverability can degrade under intensive Gaussian blur and aggressive image regeneration, especially when the attacker sacrifices data utility. Future work will incorporate adversarial training and error-bit consistency to improve evidence recovery under stronger evasion attacks.

(3) \textbf{Scaling to Low Watermark Rates and Mixed-Watermark Settings}: Multi-bit watermarking remains challenging at low watermark rates, where information density is limited. Realistic deployments may also involve multiple subsets carrying different messages, making bit-wise aggregation suboptimal. Future work will explore improved coding, noise suppression, and multi-message auditing protocols.

\section{Conclusion}
In this work, we propose \sysname, the first multi-bit dataset watermarking framework designed for auditing dataset misuse in T2I models. 
By embedding secret payloads into the latent space and using a dedicated U-Net predictor to isolate watermark residuals, \sysname enables more stable watermark propagation through the diffusion denoising process.
Experimental results show that \sysname substantially outperforms existing multi-bit baselines across datasets, model architectures, and watermark capacities.

With multi-query aggregation, \sysname achieves exact recovery at 32-bit capacity in our evaluation, improving black-box message recovery.
Combined with error-correcting codes, it further supports end-to-end tracing from recovered bits to attribution identifiers.
These results show that multi-bit auditing should go beyond bit recovery and evaluate whether watermark signals can support reliable attribution.





\section*{Acknowledgments}
We thank the anonymous reviewers for their valuable comments and suggestions.
This work was supported in part by the National Key R\&D Program of China under Grant 2025ZD0123604, the China Postdoctoral Science Foundation under Grant BX20250380, the National Natural Science Foundation of China (NSFC) under Grants 62402379, U24A20237, U22A2029, and 62672392, and the Major Science and Technology Innovation Project of Xianyang City under Grant L2025-ZDKJ-ZDCGZH-003. 
The work of Min Chen was supported in part by the CiCS project of the Gravitation research program and in part by the Dutch Research Council (NWO) under
Grant 024.006.037.
The work of Zhikun Zhang was supported in part by the NSFC under Grants 62402431 and 62441618, and in part by the Zhejiang Provincial Natural Science Foundation of China under Grant LZ26F020002.

\appendix

\bibliographystyle{ACM-Reference-Format}
\bibliography{refs_first3_and_others}
\section{Ethical Considerations}

This work studies dataset watermarking for post hoc copyright auditing of text-to-image diffusion models. 
While \sysname improves accountability and provenance, auditing results may be misinterpreted without considering statistical uncertainty.
We therefore use publicly released datasets and model checkpoints solely for non-commercial academic research under their respective licenses and do not redistribute third-party data or derived artifacts.
\sysname is intended as an auditing aid rather than an automatic enforcement mechanism, and its results should be interpreted with predefined thresholds, false-positive analysis, and supporting evidence.
Our threat model is limited to dataset usage auditing and does not target model extraction, attacks, or safety circumvention.

\section{Open Science}

To support reproducibility, we release the source code of \sysname at
\url{https://github.com/XJTU-AI-Security-Group/NotaryMark},
including implementations of watermark embedding, extraction, prediction, decoding, aggregation, as well as training and evaluation scripts and configurations.
Our experiments use only publicly available datasets and open-source diffusion models.
We provide dataset-processing scripts and model-loading instructions required to reproduce the main experiments.
No proprietary datasets, private user data, or closed-source models are required to evaluate the core claims of this work.

\section{Generative AI Usage}

During the preparation of this manuscript, we used generative AI tools, including ChatGPT, to assist with language editing and improve the clarity and organization of the writing. 
We used these tools only to support the presentation of the paper, such as refining sentences and restructuring paragraphs. 
We manually reviewed, corrected, and validated all AI-assisted text.
In addition, we used a large language model to construct a pool of 150 auditing prompts for each dataset while maintaining a linguistic style similar to that of the corresponding dataset. 
We used these prompts solely as query inputs in our black-box auditing experiments.

Generative AI tools were not used to replace the authors' scientific reasoning, experimental analysis, or conclusions.
We verified the manuscript against the implementation, experimental outputs, tables, figures, references, and technical claims. 
The authors take full responsibility for the final content of the paper.

\section{Baseline Selection and Fairness of Comparison}
\label{sec:baseline_selection}

We next clarify how the baselines in our experiments are selected and how the comparison should be interpreted. 
Our evaluation includes two groups of methods: traditional multi-bit image watermarking methods and diffusion-oriented dataset watermarking methods. 
These two groups serve different purposes in our study.

We include traditional multi-bit image watermarking methods, including \textbf{DwtDct}, \textbf{SSL}, and \textbf{RivaGAN}, not because they are specifically designed for dataset-to-model copyright auditing, but because they represent mature and widely used multi-bit watermarking paradigms in pixel, frequency, and learning-based embedding spaces. 
They provide a natural stress test for whether conventional image-level watermarks can survive the indirect ``data-to-model-to-image'' transmission channel. 
In this channel, the watermark is not directly extracted from the released watermarked images; instead, it must first be absorbed by a generative model during training or fine-tuning and then reappear in newly generated images. 
Therefore, our comparison with these methods should be understood as evaluating their transferability to dataset auditing, rather than as claiming that they are fully optimized for this setting.

Among diffusion-oriented approaches, we include \textbf{DiffusionShield} as a quantitative baseline because it provides an end-to-end watermarking mechanism for diffusion pipelines and is the closest reproducible baseline under our threat model. 
It directly optimizes watermark reconstruction alongside the diffusion training objective, making it suitable for assessing whether diffusion-aware watermark learning outperforms traditional image watermarking methods.

For \textbf{HMark}, we provide a qualitative comparison rather than a quantitative one. 
The main reason is that its public implementation and complete experimental pipeline are unavailable at the time of our evaluation. 
Faithfully reproducing \textbf{HMark} would require several non-trivial design choices, including the construction of latent spatial triggers, the LoRA fine-tuning protocol, the trigger activation strategy, and the statistical retrieval procedure. 
Different choices for these components may substantially affect watermark recovery performance, making a re-implementation unreliable as a fair baseline. 
Moreover, \textbf{HMark} focuses on associating payload bits with localized latent trigger responses, whereas our setting focuses on black-box auditing of dataset usage from naturally generated T2I outputs. 
We therefore discuss \textbf{HMark} qualitatively to clarify the technical differences, while avoiding potentially misleading quantitative results from an unofficial reproduction.

For \textbf{Luo~\etal}, we also provide a qualitative comparison instead of treating it as a separate algorithmic baseline. 
Their work proposes a workflow-oriented auditing framework that uses bit-recovery statistics as evidence for detecting unauthorized style mimicry. 
However, the watermark embedding and extraction components in that workflow are instantiated using existing image watermarking methods rather than a newly designed end-to-end multi-bit watermarking mechanism tailored to diffusion dynamics. 
Accordingly, our quantitative comparison against representative traditional image watermarking methods already covers the core technical components underlying their instantiation. 
In addition, \textbf{Luo~\etal} primarily emphasize binary auditing based on distributional recovery statistics, whereas our work targets both watermark-presence detection and high-capacity message recovery for identifier-level tracing.

The goal of our baseline evaluation is not to suggest that all compared methods are equally tailored to black-box T2I dataset auditing. 
Rather, the comparison is intended to answer three complementary questions: 
(i) whether traditional multi-bit watermarks can survive the dataset-to-model channel; 
(ii) whether diffusion-aware watermarking improves robustness in generative pipelines; and 
(iii) whether a latent-aligned design can simultaneously support reliable auditing and multi-bit tracing under black-box T2I fine-tuning.

\begin{figure*}[htbp] 
     \centering
     \begin{subfigure}[b]{0.3\textwidth}
         \centering
         \includegraphics[width=\textwidth]{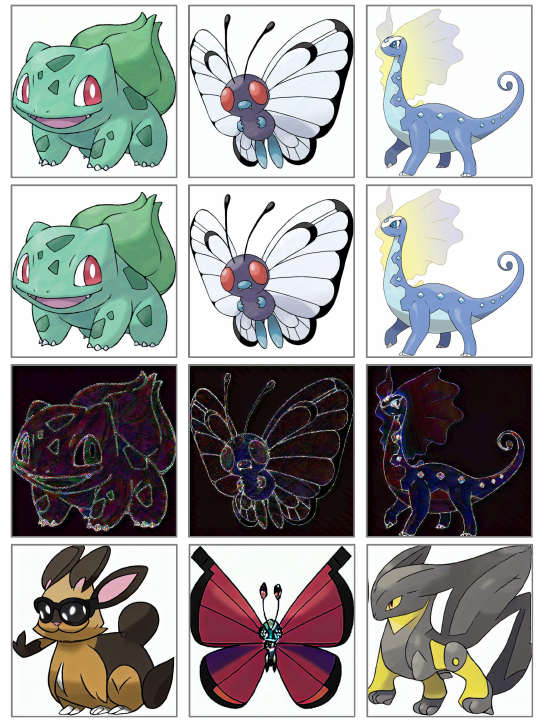}
         \caption{Pok\'{e}mon}
         \label{fig:sub-image-a}
     \end{subfigure}%
     \hfill
     \begin{subfigure}[b]{0.3\textwidth}
         \centering
         \includegraphics[width=\textwidth]{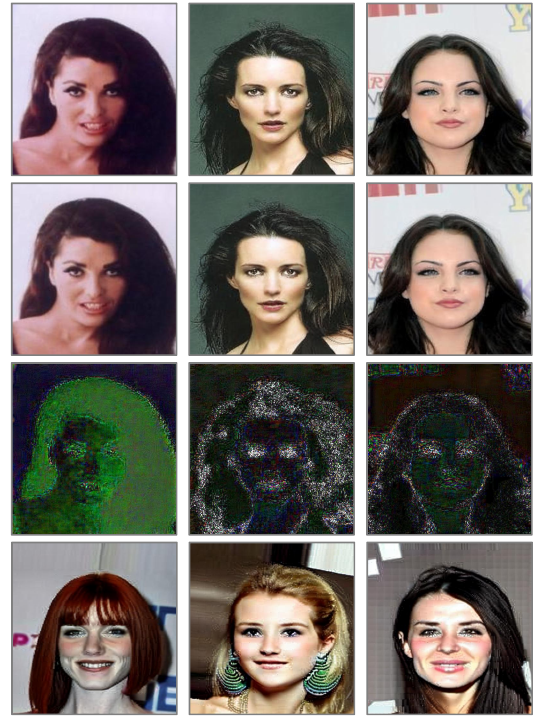}
         \caption{CelebA}
         \label{fig:sub-image-b}
     \end{subfigure}%
     \hfill
     \begin{subfigure}[b]{0.3\textwidth}
         \centering
         \includegraphics[width=\textwidth]{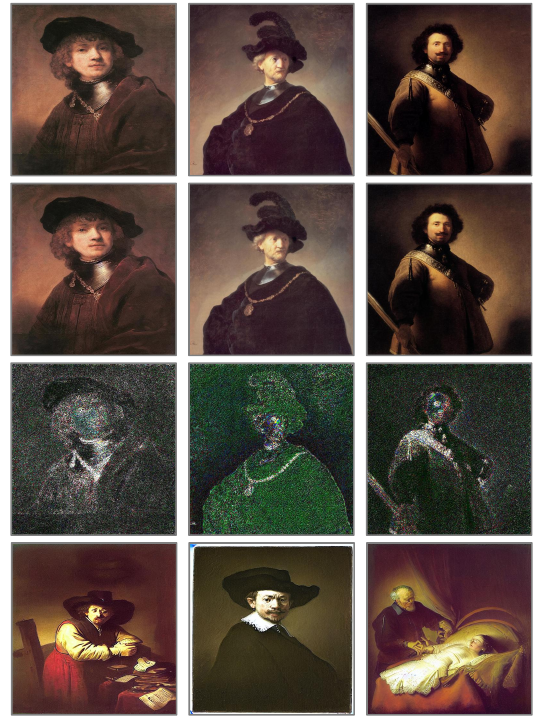}
         \caption{WikiArt}
         \label{fig:sub-image-c}
     \end{subfigure}
        
     \caption{Qualitative evaluation of \sysname on three datasets. The first row shows the original training images, followed by the corresponding images watermarked by \sysname, the pixel-wise difference maps between the watermarked and original images, and representative samples generated by the model fine-tuned on the \sysname-protected dataset.}
     \label{fig:image}%

\end{figure*} 

\section{More Details of Evaluation}
\label{sec:appendix_setup}

\subsection{Comprehensive Dataset Descriptions}
To rigorously investigate the cross-modality effectiveness of \sysname, we utilize three datasets covering distinct visual distributions:
\begin{itemize}
    \item \textbf{Pok\'{e}mon}: This collection comprises 833 high-quality images of Pok\'{e}mon creatures. Each image is paired with a descriptive caption generated by the BLIP model~\cite{pinkney2022pokemon}. It serves as a standard benchmark for evaluating fine-tuning performance within a constrained, stylized artistic domain.
    \item \textbf{CelebA}: We randomly sample 1,000 high-resolution facial images from the CelebA dataset~\cite{liu2015deep}. Moving beyond simple attribute labels, we leverage a Hugging Face image-text repository where captions were synthesized via the LLaVA model. These captions provide rich, natural-language descriptions of nuanced facial features, thereby facilitating complex text-to-image alignment.
    \item \textbf{WikiArt}: We extract 1,000 images from the Baroque style subset of the WikiArt collection~\cite{saleh2015large}. This dataset presents a significant challenge due to its intricate textures and classical composition. For consistency during fine-tuning, the prompt \textit{"A Baroque style image"} is utilized as the global identifier for this genre.
\end{itemize}
All images undergo a unified preprocessing pipeline that includes center cropping and resizing to $512 \times 512$.

\subsection{Text-to-Image Backbone Architectures}
We validate the architectural robustness of \sysname by deploying it across three diffusion backbones:
\begin{itemize}
    \item \textbf{Stable Diffusion v1.5}: A representative latent diffusion model that performs denoising within the $4 \times 64 \times 64$ latent space of a pretrained VAE. It is used as the primary baseline for evaluating multi-bit payload recovery.
    \item \textbf{Stable Diffusion XL}: A high-capacity evolution of the SD architecture featuring a $3 \times$ larger U-Net backbone and an ensemble of dual text encoders (CLIP ViT-L and OpenCLIP ViT-bigG). This model is used to assess the generalization of \sysname on large-scale parameter sets.
    \item \textbf{Kandinsky 2.2}: Unlike the SD family, Kandinsky employs an image prior module alongside latent diffusion, integrating CLIP-based image embeddings to guide the synthesis process. This allows us to verify the watermark's persistence in models with distinct conditioning mechanisms.
\end{itemize}

\subsection{Concrete Implementation Details}
\label{sec:appendix_implementation}
\mypara{Module Architectures}
\sysname contains three jointly optimized modules, including a watermark encoder $E_{\mathrm{wm}}$, a predictor $U$, and a bit decoder $D_{\mathrm{wm}}$, together with a separately trained detector $C$.
A frozen SD v1.5 AutoencoderKL maps each image $x\in\mathbb{R}^{3\times512\times512}$ to a latent $z\in\mathbb{R}^{4\times64\times64}$, while a frozen CLIP ViT-L/14 encoder provides a $768$-dimensional text condition.

$E_{\mathrm{wm}}$ first maps the $L$-bit message through a fully connected layer to a $512\times4\times4$ feature map, and then applies four transposed-convolution layers to progressively upsample it to $4\times64\times64$.
The watermarked latent is constructed as $z_w=z+\alpha w$.
$U$ is a four-level conditional U-Net with encoder channels $64$, $128$, $256$, and $512$, symmetric upsampling blocks, and skip connections, producing a residual feature $\hat{w}\in\mathbb{R}^{4\times64\times64}$.
$D_{\mathrm{wm}}$ uses four stride-$2$ convolutional blocks with channels $4\!\rightarrow\!32\!\rightarrow\!64\!\rightarrow\!128\!\rightarrow\!256$, followed by two fully connected layers, to output $L$ message logits.
$C$ adopts the same convolutional backbone, followed by adaptive average pooling and a linear classifier that outputs one watermark-presence logit.

\mypara{Optimization Configuration}
We optimize $E_{\mathrm{wm}}$, $U$, and $D_{\mathrm{wm}}$ using Adam with a learning rate of $1\times10^{-4}$, batch size $4$, zero weight decay, and no learning-rate scheduler.
During the first $\tau$ warm-up epochs, $E_{\mathrm{wm}}$ is frozen and only $U$ and $D_{\mathrm{wm}}$ are optimized using $\mathcal{L}_{\mathrm{pre}}$ and $\mathcal{L}_{\mathrm{bit}}$.
After warm-up, the three modules are jointly optimized for $100$ epochs using the full objective, with all loss weights set to $1$.
The detector $C$ is subsequently trained for $50$ epochs while the other modules remain frozen, using Adam with a learning rate of $1\times10^{-4}$, batch size $4$, zero weight decay, and binary cross-entropy over logits.

For downstream fine-tuning, we freeze the original model parameters and optimize only the LoRA adapters.
SD v1.5 is fine-tuned for $100$ epochs using AdamW with a learning rate of $1\times10^{-4}$, batch size $4$, and weight decay $10^{-2}$.
For SDXL, we insert rank-$4$ LoRA adapters into the U-Net attention projections and train them for $50$ epochs using the same learning rate, batch size, and weight decay.
For Kandinsky 2.2, we sequentially fine-tune the prior and decoder branches using rank-$4$ and rank-$32$ LoRA adapters, respectively.
Both branches use AdamW with a learning rate of $1\times10^{-5}$, batch size $1$, and zero weight decay; the prior and decoder are trained for $50$ and $50$ epochs, respectively.
All downstream models use a constant learning-rate schedule without warm-up, FP16 mixed precision, gradient accumulation of $1$, a maximum gradient norm of $1.0$, and random seed $42$.

\subsection{Statistical Significance for Auditing Decision}
\label{subsec:significance}

To provide a rigorous foundation for our auditing decision, we formulate watermark verification as a hypothesis testing problem based on the discriminator outputs, rather than relying on a fixed threshold on bit accuracy.

For each queried image, the discriminator produces a watermark-presence score $s_i \in [0,1]$, which reflects the confidence that the generated image contains the target watermark.
Given an audit trail consisting of $N$ queried images, we define the trial-level test statistic as the average detector score:
\begin{equation}
T = \frac{1}{N}\sum_{i=1}^{N} s_i.
\end{equation}

We then consider the following hypotheses:
\begin{itemize}
    \item $H_0$: the suspect model is clean and does not contain the target watermark;
    \item $H_1$: the suspect model has learned the target watermark.
\end{itemize}

To estimate the null distribution of the statistic under $H_0$, we repeatedly sample $N$ images from a clean model (or null model) and compute the corresponding trial-level statistics.
Repeating this procedure $M$ times yields an empirical null distribution:
\begin{equation}
\mathcal{T}_0 = \left\{T_j^{(0)}\right\}_{j=1}^{M}.
\end{equation}

For one observed audit trial on a suspect model, let the corresponding statistic be denoted by $T_{\mathrm{obs}}$.
We adopt a one-sided empirical hypothesis test and compute the $p$-value as:
\begin{equation}
p = \frac{\sum_{j=1}^{M}\mathbb{I}\!\left(T_j^{(0)} \geq T_{\mathrm{obs}}\right)+1}{M+1}.
\end{equation}

At significance level $\delta$, we reject the null hypothesis $H_0$ if
\begin{equation}
p < \delta,
\end{equation}
and declare the current audit trial to be positive, indicating that the suspect model contains statistically significant evidence of the target watermark.

This testing procedure provides a principled and distribution-aware decision rule for black-box auditing.
It avoids manually tuning a fixed threshold and directly controls the false positive rate via the significance level $\delta$; we use $\delta=0.005$ as the default significance level unless otherwise specified.
In our experiments, the auditing performance is summarized using TPR and FPR over repeated trials on watermarked and clean models.

\section{Experimental Results}

\subsection{Empirical False Positive Analysis}
While our experimental results yield an empirical $\text{FPR}<5\%$, we further validate the robustness of this claim by analyzing the score distribution of benign (clean) models. 
Our observations indicate that for models trained without our watermark, both the per-image bit accuracy ($\mathrm{BA}_{\text{img}}$) and the aggregated accuracy ($\mathrm{BA}_{\text{agg}}$) consistently hover around $50\%$. 

This behavior aligns perfectly with the statistical characteristics of random guessing, confirming that the extractor cannot retrieve any meaningful payload in the absence of the watermark.

\subsection{ROC Behavior in Diffusion Models}
\label{subsec:roc_analysis}

\begin{figure*}[htbp] 
     \centering
     \begin{subfigure}[b]{0.32\textwidth}
         \centering
         \includegraphics[width=\textwidth]{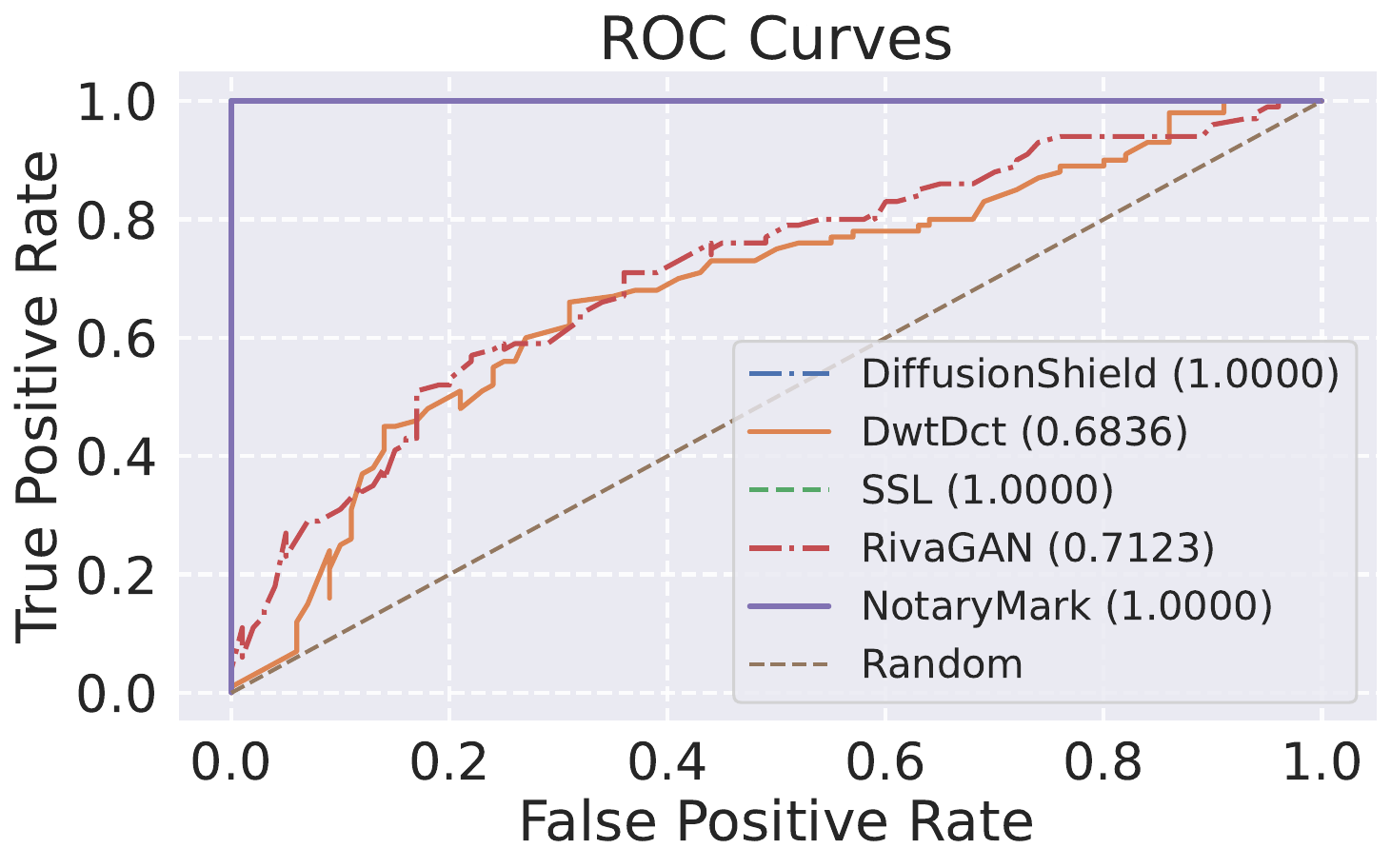}
         \caption{Pok\'{e}mon}
         \label{fig:sub-roc-a}
     \end{subfigure}%
     \hfill
     \begin{subfigure}[b]{0.32\textwidth}
         \centering
         \includegraphics[width=\textwidth]{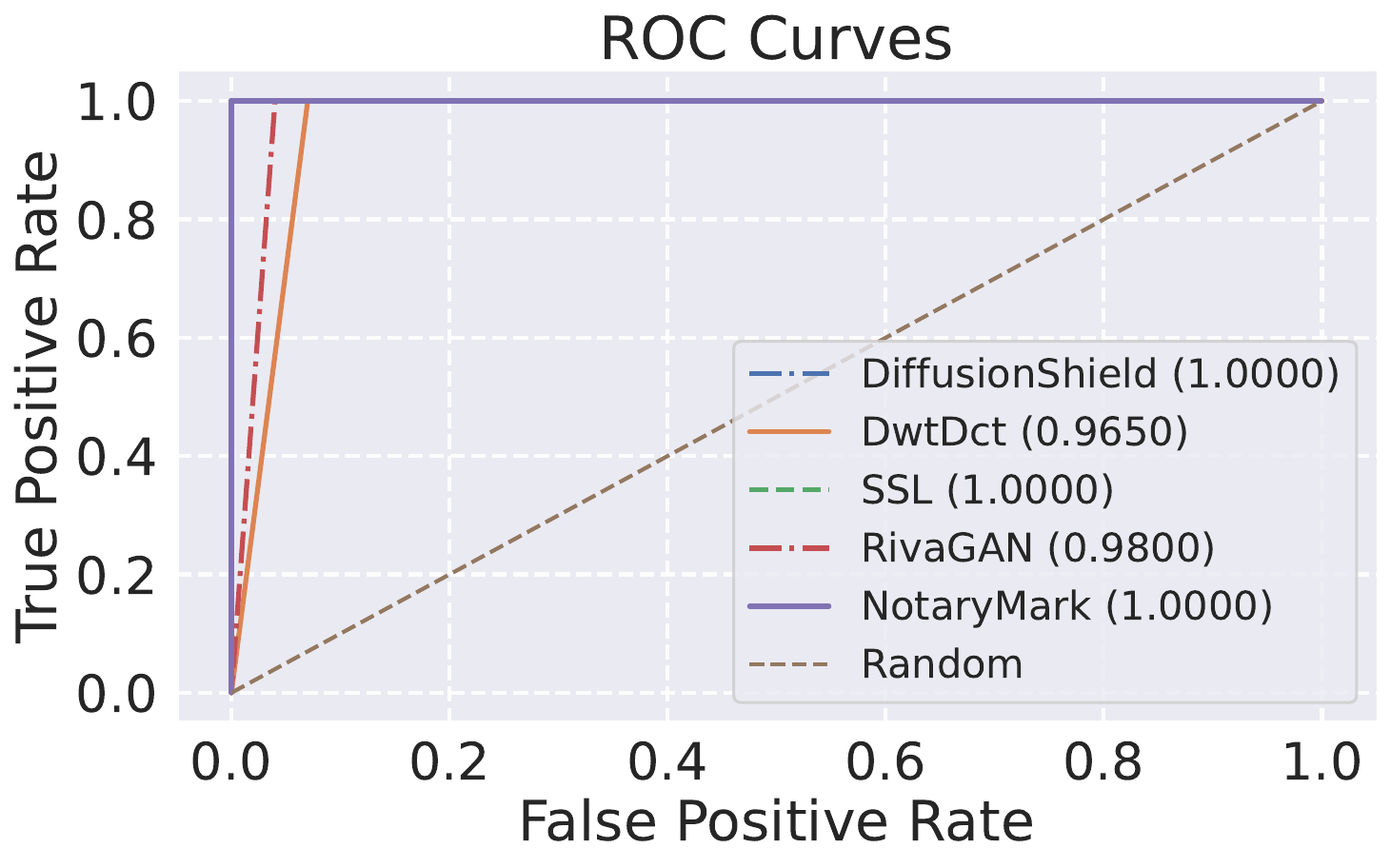}
         \caption{CelebA}
         \label{fig:sub-roc-b}
     \end{subfigure}%
     \hfill
     \begin{subfigure}[b]{0.32\textwidth}
         \centering
         \includegraphics[width=\textwidth]{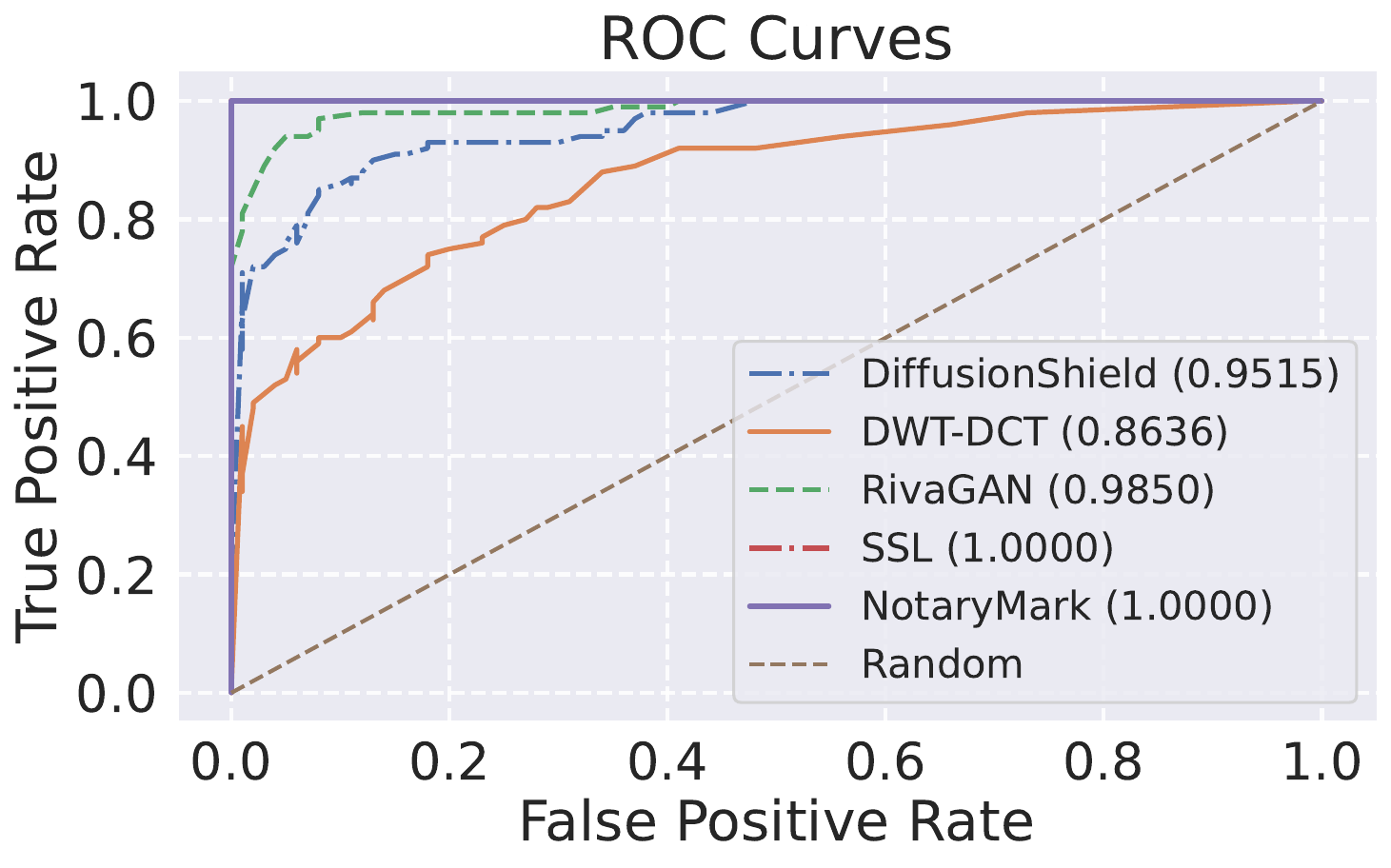}
         \caption{WikiArt}
         \label{fig:sub-roc-c}
     \end{subfigure}
        
     \caption{ROC curve comparison of different methods on three datasets.
     Each subfigure illustrates the trade-off between the True Positive Rate (TPR) and the False Positive Rate (FPR).}
     \label{fig:roc}
\end{figure*}

\autoref{fig:roc} presents the ROC curves of \sysname and representative baseline methods after diffusion fine-tuning on three datasets.
The ROC curves are obtained by varying the significance level $\delta$ in the hypothesis test described in~\autoref{subsec:significance}.
For each value of $\delta$, an audit trial is declared positive when its empirical $p$-value satisfies $p < \delta$.
We then compute the corresponding TPR and FPR over repeated trials on watermarked and clean models, respectively.
By sweeping $\delta$, we obtain a sequence of operating points and use them to construct the ROC curve.

Overall, the ROC results show that verification behavior in diffusion models is highly method- and data-dependent.
Among all compared methods, \sysname exhibits the most stable and ideal behavior, consistently achieving an AUC of 1.0 across all three datasets.

On Pok\'{e}mon, the performance gap between methods is particularly pronounced: although DiffusionShield and SSL also achieve AUC = 1.0, DwtDct and RivaGAN are substantially weaker, with AUC values of 0.6836 and 0.7123, respectively. 
On CelebA, most methods exhibit strong ranking ability, with DwtDct and RivaGAN reaching 0.9650 and 0.9800, while DiffusionShield, SSL, and \sysname all attain 1.0. 
On WikiArt, the differences become clearer again: DwtDct yields the weakest result (0.8636), DiffusionShield and RivaGAN achieve 0.9515 and 0.9850, and both SSL and \sysname reach 1.0. 
Overall, \sysname maintains the most robust and consistent verification capability across datasets with diverse visual characteristics.

Importantly, strong ROC behavior should not be conflated with accurate multi-bit payload recovery.
A method may still achieve a high AUC by capturing coarse watermark-presence signatures that are sufficient for binary discrimination, while failing to recover the embedded message with high fidelity.
This distinction is particularly important for dataset auditing, where practical attribution requires not only reliable auditing decisions but also accurate identifier extraction for downstream tracing.
From this perspective, the main advantage of \sysname is that it provides both: highly separable verification statistics for auditing and sufficiently stable multi-bit signals for precise attribution.

\subsection{Qualitative Results}
\label{subsec:qualitative_analysis}
\autoref{fig:image} provides qualitative comparisons between clean and watermarked samples generated under the same prompts.
Across all datasets, including Pok\'{e}mon, CelebA, and WikiArt, the two sets of images appear visually indistinguishable. 
There are no noticeable artifacts, color shifts, or texture anomalies that often plague traditional pixel-domain watermarking methods. 
Even in regions with high-frequency details, such as human hair textures or intricate character outlines, the structural integrity remains perfectly intact.

We further observe that fine-tuning on protected data preserves the intended semantic content and artistic style without inducing visible style drift or fidelity degradation. 
\sysname maintains the sharp contrast and vibrant palette characteristic of the original diffusion models, suggesting that the aligned latent denoising dynamics successfully prevent the watermark from interfering with the primary generative task. 

To rigorously assess imperceptibility, we also examine the residual maps (\ie, the amplified difference between the clean and watermarked images). 
Residual map analysis reveals that the watermark energy does not diffuse randomly as stochastic noise; instead, it exhibits a high degree of content adaptivity, in which the watermark signals are tightly coupled to the image's underlying textures and structural features. 
By achieving structural alignment with the image content in the latent space, this intrinsic correlation ensures that attackers cannot readily remove the watermark, as they would with conventional random noise.
Any attempt to forcibly remove the watermark through denoising or smoothing would inevitably compromise the image's structural integrity and textural quality. 
This deep integration not only ensures high visual imperceptibility but also strengthens the robustness of \sysname against malicious tampering and black-box auditing attacks.



\begin{table*}[t]
\centering
\caption{Ablation study of \sysname with different loss combinations.}
\label{tab:ablation}
\footnotesize
\setlength{\tabcolsep}{4.2pt}
\renewcommand{\arraystretch}{0.9}
\begin{tabular}{l|cccc|cc}
\toprule
Loss terms enabled & TPR (\%)$\uparrow$ & FPR (\%)$\downarrow$ & $BA_{img}$ (\%)$\uparrow$ & $BA_{agg}$ (\%)$\uparrow$ & LPIPS$\downarrow$ & FID$\downarrow$ \\
\midrule
$\mathcal{L}_{\text{bit}}$
& 100.00 & 0.00 & 73.94 & 96.13 & 0.0188 & 89.07 \\

$\mathcal{L}_{\text{bit}} + \mathcal{L}_{\text{pre}}$
& 100.00 & 0.00 & 77.42 & 98.94 & 0.0122 & 104.68 \\

$\mathcal{L}_{\text{bit}} + \mathcal{L}_{\text{pre}} + \mathcal{L}_{\text{imp}}$
& 100.00 & 0.00 & 85.35 & 100.00 & 0.0112 & 77.23 \\
\bottomrule
\end{tabular}
\end{table*}

\subsection{Joint Effects of Watermark Rate and Strength}
\label{subsec:joint_rate_strength}

We further examine the joint effects of watermark strength $\alpha$ and watermark rate $\rho$ by varying both parameters from 0.2 to 1.0 on the three datasets.
\autoref{tab:alpha_rho_pokemon}, \autoref{tab:alpha_rho_celeba}, and \autoref{tab:alpha_rho_wikiart} report the resulting auditing performance, message recovery, and generation quality.
It should be noted that FID measures the distributional distance between images generated by the clean fine-tuned model and those generated by the corresponding watermarked fine-tuned model under the same prompts.

Overall, increasing either watermark strength or watermark rate improves watermark recoverability, while the two parameters can partially compensate for each other.
When both parameters are small, the watermark signal is insufficiently represented during fine-tuning.
For example, at $(\alpha,\rho)=(0.2,0.2)$, $\mathrm{BA}_{\text{agg}}$ remains close to random guessing on all three datasets, while the TPR values are only 79.00\%, 0.00\%, and 52.00\% on Pok\'emon, CelebA, and WikiArt, respectively.
At a fixed moderate strength of $\alpha=0.4$, increasing $\rho$ from 0.2 to 1.0 raises $\mathrm{BA}_{\text{agg}}$ from 50.31\% to 97.46\% on Pok\'emon, from 57.47\% to 90.75\% on CelebA, and from 57.53\% to 90.47\% on WikiArt.
This confirms that exposing the downstream model to more protected samples substantially strengthens the accumulated watermark evidence.

The results also reveal that reliable auditing decisions are easier to achieve than exact multi-bit message recovery.
In several low-budget configurations, the TPR already approaches or reaches 100\%, while $\mathrm{BA}_{\text{agg}}$ remains relatively low.
For instance, at $(\alpha,\rho)=(0.4,0.2)$, the TPR reaches 100.00\% on Pok\'emon and WikiArt and 97.00\% on CelebA, whereas the corresponding $\mathrm{BA}_{\text{agg}}$ values are only 50.31\%, 57.53\%, and 57.47\%.
Once $\alpha\geq0.6$ and $\rho\geq0.4$, however, all three datasets achieve a TPR of 100.00\%, and $\mathrm{BA}_{\text{agg}}$ remains above 92.78\%.
This indicates that moderate parameter settings are sufficient for reliable identifier-level recovery across different data domains.
Therefore, practical configurations should use a moderate $\alpha$ together with a sufficiently large $\rho$, rather than maximizing both parameters.
\begin{DIFnomarkup}
\begin{table}[t]
\centering
\caption{Performance of \sysname\ under different combinations of $\alpha$ and $\rho$ on the Pok\'emon dataset. The first and second values in each parameter combination correspond to $\alpha$ and $\rho$, respectively.}
\label{tab:alpha_rho_pokemon}
\scriptsize
\setlength{\tabcolsep}{3.0pt}
\renewcommand{\arraystretch}{0.9}

\begin{minipage}[t]{0.485\linewidth}
\centering
\begin{tabular*}{\linewidth}{
    @{\extracolsep{\fill}}c|ccccc@{}
}
\toprule
\multicolumn{6}{c}{\textbf{TPR (\%) $\uparrow$}} \\
\midrule
$\alpha \backslash \rho$
& 0.2 & 0.4 & 0.6 & 0.8 & 1.0 \\
\midrule
0.2 & 79.00  & 83.00  & 94.00  & 100.00 & 100.00 \\
0.4 & 100.00 & 100.00 & 100.00 & 100.00 & 100.00 \\
0.6 & 100.00 & 100.00 & 100.00 & 100.00 & 100.00 \\
0.8 & 100.00 & 100.00 & 100.00 & 100.00 & 100.00 \\
1.0 & 100.00 & 100.00 & 100.00 & 100.00 & 100.00 \\
\bottomrule
\end{tabular*}
\end{minipage}
\hfill
\begin{minipage}[t]{0.485\linewidth}
\centering
\begin{tabular*}{\linewidth}{
    @{\extracolsep{\fill}}c|ccccc@{}
}
\toprule
\multicolumn{6}{c}{\textbf{$BA_{img}$ (\%) $\uparrow$}} \\
\midrule
$\alpha \backslash \rho$
& 0.2 & 0.4 & 0.6 & 0.8 & 1.0 \\
\midrule
0.2 & 50.25 & 50.31 & 50.62 & 52.92 & 62.19 \\
0.4 & 50.10 & 57.98 & 66.46 & 69.92 & 81.98 \\
0.6 & 51.21 & 65.87 & 72.00 & 78.90 & 84.25 \\
0.8 & 54.37 & 67.50 & 74.98 & 79.94 & 89.83 \\
1.0 & 56.35 & 72.00 & 84.27 & 89.94 & 90.69 \\
\bottomrule
\end{tabular*}
\end{minipage}

\vspace{8pt}

\begin{minipage}[t]{0.485\linewidth}
\centering
\begin{tabular*}{\linewidth}{
    @{\extracolsep{\fill}}c|ccccc@{}
}
\toprule
\multicolumn{6}{c}{\textbf{$BA_{agg}$ (\%) $\uparrow$}} \\
\midrule
$\alpha \backslash \rho$
& 0.2 & 0.4 & 0.6 & 0.8 & 1.0 \\
\midrule
0.2 & 51.75 & 51.50 & 53.12 & 58.87 & 84.78 \\
0.4 & 50.31 & 75.15 & 94.15 & 94.87 & 97.46 \\
0.6 & 56.00 & 94.16 & 95.62 & 98.65 & 100.00 \\
0.8 & 67.66 & 95.56 & 99.22 & 99.59 & 100.00 \\
1.0 & 71.34 & 97.72 & 100.00 & 99.94 & 100.00 \\
\bottomrule
\end{tabular*}
\end{minipage}
\hfill
\begin{minipage}[t]{0.485\linewidth}
\centering
\begin{tabular*}{\linewidth}{
    @{\extracolsep{\fill}}c|ccccc@{}
}
\toprule
\multicolumn{6}{c}{\textbf{FID $\downarrow$}} \\
\midrule
$\alpha \backslash \rho$
& 0.2 & 0.4 & 0.6 & 0.8 & 1.0 \\
\midrule
0.2 & 90.43 & 89.40 & 89.39 & 88.39 & 91.51 \\
0.4 & 93.41 & 91.13 & 89.87 & 89.22 & 93.69 \\
0.6 & 88.04 & 92.39 & 92.46 & 91.46 & 88.90 \\
0.8 & 92.20 & 90.91 & 89.28 & 91.25 & 94.47 \\
1.0 & 90.35 & 92.23 & 95.32 & 96.37 & 102.00 \\
\bottomrule
\end{tabular*}
\end{minipage}

\end{table}
\end{DIFnomarkup}

\begin{DIFnomarkup}
\begin{table}[t]
\centering
\caption{Performance of \sysname\ under different combinations of $\alpha$ and $\rho$ on the CelebA dataset.}
\label{tab:alpha_rho_celeba}
\scriptsize
\setlength{\tabcolsep}{3.0pt}
\renewcommand{\arraystretch}{0.9}

\begin{minipage}[t]{0.485\linewidth}
\centering
\begin{tabular*}{\linewidth}{
    @{\extracolsep{\fill}}c|ccccc@{}
}
\toprule
\multicolumn{6}{c}{\textbf{TPR (\%) $\uparrow$}} \\
\midrule
$\alpha \backslash \rho$
& 0.2 & 0.4 & 0.6 & 0.8 & 1.0 \\
\midrule
0.2 & 0.00  & 0.00   & 0.00   & 33.00  & 100.00 \\
0.4 & 97.00 & 100.00 & 100.00 & 100.00 & 100.00 \\
0.6 & 77.00 & 100.00 & 100.00 & 100.00 & 100.00 \\
0.8 & 100.00 & 100.00 & 100.00 & 100.00 & 100.00 \\
1.0 & 99.00 & 100.00 & 100.00 & 100.00 & 100.00 \\
\bottomrule
\end{tabular*}
\end{minipage}
\hfill
\begin{minipage}[t]{0.485\linewidth}
\centering
\begin{tabular*}{\linewidth}{
    @{\extracolsep{\fill}}c|ccccc@{}
}
\toprule
\multicolumn{6}{c}{\textbf{$BA_{img}$ (\%) $\uparrow$}} \\
\midrule
$\alpha \backslash \rho$
& 0.2 & 0.4 & 0.6 & 0.8 & 1.0 \\
\midrule
0.2 & 50.27 & 51.81 & 51.98 & 56.13 & 72.29 \\
0.4 & 51.52 & 54.10 & 57.35 & 62.37 & 73.37 \\
0.6 & 51.15 & 72.83 & 82.31 & 87.48 & 94.81 \\
0.8 & 52.40 & 83.58 & 88.81 & 93.81 & 97.90 \\
1.0 & 54.96 & 87.92 & 92.08 & 96.21 & 98.73 \\
\bottomrule
\end{tabular*}
\end{minipage}

\vspace{8pt}

\begin{minipage}[t]{0.485\linewidth}
\centering
\begin{tabular*}{\linewidth}{
    @{\extracolsep{\fill}}c|ccccc@{}
}
\toprule
\multicolumn{6}{c}{\textbf{$BA_{agg}$ (\%) $\uparrow$}} \\
\midrule
$\alpha \backslash \rho$
& 0.2 & 0.4 & 0.6 & 0.8 & 1.0 \\
\midrule
0.2 & 51.50 & 57.72 & 58.34 & 71.28 & 80.22 \\
0.4 & 57.47 & 62.63 & 64.31 & 83.41 & 90.75 \\
0.6 & 55.66 & 92.78 & 94.78 & 97.53 & 100.00 \\
0.8 & 59.78 & 95.66 & 99.16 & 99.94 & 100.00 \\
1.0 & 70.78 & 99.47 & 96.19 & 99.00 & 100.00 \\
\bottomrule
\end{tabular*}
\end{minipage}
\hfill
\begin{minipage}[t]{0.485\linewidth}
\centering
\begin{tabular*}{\linewidth}{
    @{\extracolsep{\fill}}c|ccccc@{}
}
\toprule
\multicolumn{6}{c}{\textbf{FID $\downarrow$}} \\
\midrule
$\alpha \backslash \rho$
& 0.2 & 0.4 & 0.6 & 0.8 & 1.0 \\
\midrule
0.2 & 56.65  & 62.90  & 68.29  & 71.54  & 78.05 \\
0.4 & 59.86  & 62.60  & 72.59  & 80.78  & 83.05 \\
0.6 & 76.58  & 89.96  & 90.77  & 94.95  & 101.11 \\
0.8 & 85.81  & 108.72 & 113.97 & 138.00 & 149.31 \\
1.0 & 113.58 & 118.47 & 132.41 & 153.91 & 156.95 \\
\bottomrule
\end{tabular*}
\end{minipage}

\end{table}
\end{DIFnomarkup}

\begin{DIFnomarkup}
\begin{table}[t]
\centering
\caption{Performance of \sysname\ under different combinations of $\alpha$ and $\rho$ on the WikiArt dataset.}
\label{tab:alpha_rho_wikiart}
\scriptsize
\setlength{\tabcolsep}{3.0pt}
\renewcommand{\arraystretch}{0.9}

\begin{minipage}[t]{0.485\linewidth}
\centering
\begin{tabular*}{\linewidth}{
    @{\extracolsep{\fill}}c|ccccc@{}
}
\toprule
\multicolumn{6}{c}{\textbf{TPR (\%) $\uparrow$}} \\
\midrule
$\alpha \backslash \rho$
& 0.2 & 0.4 & 0.6 & 0.8 & 1.0 \\
\midrule
0.2 & 52.00  & 100.00 & 100.00 & 100.00 & 100.00 \\
0.4 & 100.00 & 100.00 & 100.00 & 100.00 & 100.00 \\
0.6 & 100.00 & 100.00 & 100.00 & 100.00 & 100.00 \\
0.8 & 100.00 & 100.00 & 100.00 & 100.00 & 100.00 \\
1.0 & 100.00 & 100.00 & 100.00 & 100.00 & 100.00 \\
\bottomrule
\end{tabular*}
\end{minipage}
\hfill
\begin{minipage}[t]{0.485\linewidth}
\centering
\begin{tabular*}{\linewidth}{
    @{\extracolsep{\fill}}c|ccccc@{}
}
\toprule
\multicolumn{6}{c}{\textbf{$BA_{img}$ (\%) $\uparrow$}} \\
\midrule
$\alpha \backslash \rho$
& 0.2 & 0.4 & 0.6 & 0.8 & 1.0 \\
\midrule
0.2 & 49.67 & 50.58 & 51.21 & 52.08 & 74.88 \\
0.4 & 52.08 & 54.30 & 55.60 & 62.00 & 79.29 \\
0.6 & 61.77 & 86.54 & 92.48 & 94.73 & 97.83 \\
0.8 & 76.04 & 94.69 & 95.17 & 93.29 & 97.60 \\
1.0 & 70.98 & 94.75 & 97.12 & 95.21 & 99.23 \\
\bottomrule
\end{tabular*}
\end{minipage}

\vspace{8pt}

\begin{minipage}[t]{0.485\linewidth}
\centering
\begin{tabular*}{\linewidth}{
    @{\extracolsep{\fill}}c|ccccc@{}
}
\toprule
\multicolumn{6}{c}{\textbf{$BA_{agg}$ (\%) $\uparrow$}} \\
\midrule
$\alpha \backslash \rho$
& 0.2 & 0.4 & 0.6 & 0.8 & 1.0 \\
\midrule
0.2 & 50.59 & 54.00 & 56.47 & 60.03 & 87.50 \\
0.4 & 57.53 & 58.16 & 58.73 & 82.34 & 90.47 \\
0.6 & 90.78 & 100.00 & 100.00 & 100.00 & 100.00 \\
0.8 & 99.84 & 100.00 & 100.00 & 100.00 & 100.00 \\
1.0 & 99.28 & 100.00 & 100.00 & 100.00 & 100.00 \\
\bottomrule
\end{tabular*}
\end{minipage}
\hfill
\begin{minipage}[t]{0.485\linewidth}
\centering
\begin{tabular*}{\linewidth}{
    @{\extracolsep{\fill}}c|ccccc@{}
}
\toprule
\multicolumn{6}{c}{\textbf{FID $\downarrow$}} \\
\midrule
$\alpha \backslash \rho$
& 0.2 & 0.4 & 0.6 & 0.8 & 1.0 \\
\midrule
0.2 & 76.79  & 89.99  & 92.81  & 100.39 & 109.12 \\
0.4 & 86.30  & 90.19  & 95.30  & 105.58 & 104.03 \\
0.6 & 119.58 & 133.78 & 124.53 & 114.72 & 124.73 \\
0.8 & 168.85 & 201.41 & 235.11 & 232.31 & 237.19 \\
1.0 & 157.85 & 188.14 & 214.72 & 237.05 & 246.81 \\
\bottomrule
\end{tabular*}
\end{minipage}

\end{table}
\end{DIFnomarkup}

\subsection{Computational Efficiency and Deployment Feasibility}
To evaluate the practical feasibility of \sysname in real-world production environments, we quantify the training overhead and inference latency of its core components, including the watermark encoder, predictor, and decoder.
Our experiments take place on a workstation equipped with four NVIDIA GeForce RTX 3090 (24GB VRAM) GPUs. 
Benefiting from a lightweight architectural design, the joint training process of the entire watermarking system exhibits exceptional convergence speed.
On the Pok\'{e}mon dataset, a complete training cycle finishes within 3 to 5 hours.
Furthermore, the inference latency for embedding a watermark into a single image is only 0.3 seconds. 
Such low computational overhead ensures that \sysname operates efficiently and maintains good performance in AIGC generation tasks.

\section{Ablation Study}

\subsection{Quantifying the Contributions of Progressive Loss Components}
We design a series of progressive experimental configurations to systematically evaluate the contribution of each loss component in \sysname. 
The ablation begins with the core verification logic, employing only the fundamental bit decoding loss $\mathcal{L}_{\text{bit}}$ to establish a baseline for extraction capability. 
To mitigate the inherent stochastic interference during diffusion sampling, we subsequently introduce the predictability loss, $\mathcal{L}_{\text{pre}}$, to create an intermediate configuration that enhances watermark recovery robustness. 
Finally, to meet the stringent fidelity requirements of practical applications, we integrate the full objective function, including the imperceptibility regularizer $\mathcal{L}_{\text{imp}}$, to enforce explicit constraints on generation quality across both pixel and latent spaces. 

\autoref{tab:ablation} reports the audit performance and generation quality under different loss combinations.
The results show that $\mathcal{L}_{\text{bit}}$ alone already establishes reliable auditability, achieving TPR/FPR of 100.00\%/0.00\% with $BA_{img}=73.94\%$ and $BA_{agg}=96.13\%$.
Adding the predictability loss $\mathcal{L}_{\text{pre}}$ further improves watermark recoverability, increasing $BA_{img}$ to 77.42\% and $BA_{agg}$ to 98.94\%.
These results suggest that enforcing predictable latent residual patterns helps the model recover watermark signals more accurately and consistently.

However, the combination of $\mathcal{L}_{\text{bit}}$ and $\mathcal{L}_{\text{pre}}$ alone still does not yield the best overall trade-off.
FID rises from 89.07 to 104.68, indicating degraded model-level generation quality.
To address this issue, we further incorporate the imperceptibility regularizer $\mathcal{L}_{\text{imp}}$, which jointly constrains pixel-level distortion and latent-space deviation.
With the full objective, \sysname achieves the best overall performance: $BA_{img}$ increases to 85.35\%, $BA_{agg}$ reaches 100.00\%, and both LPIPS and FID are improved to 0.0112 and 77.23, respectively.

These results indicate that $\mathcal{L}_{\text{bit}}$ is sufficient to establish reliable auditing, $\mathcal{L}_{\text{pre}}$ further enhances watermark recoverability, and $\mathcal{L}_{\text{imp}}$ is critical for jointly optimizing recovery accuracy and generation quality.

\begin{table}[ht]
\centering
\caption{Audit performance across different numbers of query prompts. The metrics are reported based on $\mathrm{BA}_{\text{agg}} (\%)$  across multiple experimental runs.}
\label{tab:prompt_efficiency}
\footnotesize
\setlength{\tabcolsep}{7.0pt}
\renewcommand{\arraystretch}{0.9}
\begin{tabular}{c|cccc}
\toprule
\textbf{Prompts ($N$)} & \textbf{Mean $\uparrow$} & \textbf{Std$\downarrow$} & \textbf{Min} & \textbf{Max} \\ 
\midrule
5  & 96.59  & 4.90 & 68.75  & 100.00 \\
10 & 98.59  & 2.27 & 87.50  & 100.00 \\
15 & 99.78  & 0.79 & 96.88  & 100.00 \\
20 & \textbf{100.00} & \textbf{0.00} & 100.00 & 100.00 \\
25 & 99.94  & 0.44 & 96.88  & 100.00 \\
30 & \textbf{100.00} & \textbf{0.00} & 100.00 & 100.00 \\
35 & \textbf{100.00} & \textbf{0.00} & 100.00 & 100.00 \\
\bottomrule
\end{tabular}
\end{table}

\subsection{Impact of Query Budget on Auditing Certainty}
To evaluate the practicality of \sysname in resource-constrained scenarios, we conduct an ablation study on the query budget, defined as the number of prompts $N$ used for auditing.
The query budget is a critical factor determining the computational cost and latency of the auditing process.
As summarized in~\autoref{tab:prompt_efficiency}, \sysname demonstrates exceptional sample efficiency. 
Even with an extremely limited budget of $N=5$, the system achieves a mean bit accuracy of $96.59\%$. 
As the budget increases to $N=15$, the $BA_{agg}$ rises to $99.78\%$ while the standard deviation (Std) drops significantly, indicating enhanced stability in the verification signal.

Notably, the auditing performance consistently reaches a plateau of $100.00\%$ accuracy when $N \geq 30$. 
These results suggest that an auditor can achieve near-perfect certainty with minimal computational overhead (\eg, generating 30 images in $\approx 60$ seconds). 
Such efficiency positions \sysname as a highly scalable solution for automated, high-throughput model screening in real-world regulatory environments.

\end{document}